\documentclass[ba]{imsart}
\pubyear{2021}
\volume{TBA}
\issue{TBA}
\firstpage{1}
\lastpage{1}

\usepackage{amsthm}
\usepackage{amsmath}
\usepackage{natbib}
\usepackage[colorlinks,citecolor=blue,urlcolor=blue,filecolor=blue,backref=page]{hyperref}
\usepackage{graphicx}
\usepackage{dblfloatfix}

\usepackage{float}
\usepackage{natbib}
\usepackage{xcolor}
\usepackage{hyperref}
\usepackage{caption}
\usepackage{subcaption}
\startlocaldefs
\endlocaldefs

\begin{document}


\begin{frontmatter}
\title{Visualizations for Bayesian Additive Regression Trees\support{Alan Inglis and Andrew Parnell's work was supported by a Science Foundation Ireland Career Development Award grant 17/CDA/4695. In addition Andrew Parnell’s work was supported by: an investigator award (16/IA/4520); a Marine Research Programme funded by the Irish Government, co-financed by the European Regional Development Fund (Grant-Aid Agreement No. PBA/CC/18/01); European Union’s Horizon 2020 research and innovation programme InnoVar under grant agreement No 818144; SFI Centre for Research Training in Foundations of Data Science 18CRT/6049, and SFI Research Centre awards I-Form 16/RC/3872 and Insight 12/RC/2289\_P2. For the purpose of Open Access, the author has applied a CC BY public copyright licence to any Author Accepted Manuscript version arising from this submission.}}
\runtitle{}

\begin{aug}
\author{\fnms{Alan} \snm{Inglis}\thanksref{addr1}\ead[label=e1]{alan.inglis@mu.ie}},
\author{\fnms{Andrew} \snm{Parnell}\thanksref{addr2, t1}\ead[label=e2]{andrew.parnell@mu.ie}},
\and
\author{\fnms{Catherine} \snm{Hurley}\thanksref{addr3, t1}\ead[label=e3]{catherine.hurley@mu.ie}}

\runauthor{}

\address[addr1]{Hamilton Institute, Maynooth University 
    \printead{e1}
}
\address[addr2]{Hamilton Institute, Insight Centre for Data Analytics, Maynooth University 
    \printead{e2}
}
\address[addr3]{Department of Mathematics and Statistics, Maynooth University 
    \printead{e3}
}


\end{aug}

\begin{abstract}
Tree-based regression and classification has become a standard tool in modern data science. Bayesian Additive Regression Trees (BART) has in particular gained wide popularity due its flexibility in dealing with interactions and non-linear effects. BART is a Bayesian tree-based machine learning method that can be applied to both regression and classification problems and yields competitive or superior results when compared to other predictive models. As a Bayesian model, BART allows the practitioner to explore the uncertainty around predictions through the posterior distribution. In this paper, we present new visualization techniques for exploring BART models. We construct conventional plots to analyze a model’s performance and stability as well as create new tree-based plots to analyze variable importance, interaction, and tree structure. We employ Value Suppressing Uncertainty Palettes (VSUP) to construct heatmaps that display variable importance and interactions jointly  using  color scale to represent posterior uncertainty. Our new visualizations are designed to work with the most popular BART R packages available, namely \texttt{BART}, \texttt{dbarts}, and \texttt{bartMachine}. Our approach is implemented in the  R package \texttt{bartMan} (BART Model ANalysis).
\end{abstract}


\end{frontmatter}

\section{Introduction}
\label{sec:intro}

Bayesian Additive Regression Trees \citep[BART;][]{chipman2010bart} is a non-parametric sum-of-trees-based ensemble method. BART has been shown to be a useful predictive tool and has been applied in diverse areas such as risk management \citep{liu2015ensemble}, proteomics \citep{hernandez2015bayesian}, and avalanche forecasting \citep{blattenberger2014avalanche}. The BART method has also been extended into many areas, such as survival analysis \citep{sparapani2016nonparametric} and causal inference \citep{hill2011bayesian, hahn2020bayesian}. Its excellent empirical performance has motivated works on its theoretical foundations \citep{lineroAnDyang2018, prado2021bayesian}. 
BART now enjoys widespread use due to its competitive performance against other tree-based predictive models, such as Random Forest \citep{breiman2001randomforest} and Gradient Boosted Trees 
\citep{FriedmanGBM}.

BART models are used for making predictions for both binary and continuous response variables and are fit using the \texttt{R} packages \texttt{dbarts} \citep{DBARTpack}, \texttt{bartMachine} \citep{BARTMachPack}, and \texttt{BART} \citep{BARTpack}, among others. These packages offer limited visualizations  and in some cases leave it to the user to manually create their own visualizations by extracting information from the fitted model. Our goal is to create novel visualizations and to streamline this process for the aforementioned BART packages by creating a suite of plots for visualizing and evaluating both the BART fit  and the posterior distribution. In our work, various aspects of a BART model can be assessed (e.g., variable importance and variable interaction) by analyzing the structure of the trees used in the model. However, our approach goes beyond many standard machine learning visualization techniques by allowing for uncertainty in the posterior to propagate into the diagrams.

One of the more challenging aspects of model visualization is the depiction of uncertainty. The predictions from the BART models we create exhibit uncertainty associated with the posterior distribution, and the way we choose to represent this uncertainty may have an impact on how the model is analyzed and how our audience interprets the findings. 
This issue has been well studied in areas that regularly deal with uncertainties in data \citep[e.g. ][]{uncert1,uncert2}. For example, error bars, confidence intervals, or quantile intervals are common tools used to display uncertainty. However, these tools cannot be universally applied to all situations where displaying the uncertainty is necessary, such as in heatmaps or point clouds. When using point clouds to map data over many iterations, 95\% confidence ellipses can be used to encircle points. An example of this can be seen in Section \ref{sec:mds}.

Methods for producing visualizations of importance and interaction for standard machine learning models can be found in \cite{inglis2022visualizing}. However, in Bayesian models it is important to include the uncertainties that arise as part of the calculation of a full joint posterior distribution. Our new displays use a method called Value Suppressing Uncertainty Palettes \citep{vsup}, which allows for both the value and the uncertainty to be displayed in a single plot. Traditional methods for displaying a value and uncertainty simultaneously require a 2D bivariate map, conventionally displayed as a square \citep[for example, see ][]{robertson1986generation,teuling2011bivariate}. However, due to the large color-space of 2D bivariate maps, the ability to distinguish between two different visual aspects can become challenging. VSUPs improve on this method by using an arc to assign colors and blend together data values with high uncertainty so that values become more distinguishable as the uncertainty decreases. This reduction of the visual color-space helps to both distinguish between low and high uncertainty and promotes caution when the uncertainty is high \citep{vsup}. 

By examining the trees in a BART model we can learn about the stability and variability of tree structures as the algorithm iterates to build the posterior. We offer new tree-based plots that focus attention on certain aspects of the model fit in an intuitive way. We provide space-saving layouts as well as providing various sorting/filtering methods and coloring options. When combined with ordination techniques, we provide easy to use tools which aid in highlighting interesting aspects of the model fit, such as variable importance or common interactions. 

Multidimensional scaling (MDS) plots  are a common method for graphically displaying relationships between objects in multidimensional space \citep{torgerson1952multidimensional}. Objects that are similar appear closer on the graph, whereas objects objects that are less similar are farther away. MDS can be used to reduce the number of dimensions in high-dimensional data as well as interpret dissimilarities as graph distances. We construct an MDS display of a BART fit and extend it to display the uncertainty. For each iteration of the BART fit, we perform MDS on proximities and rotate each plot to match a particular target iteration. From this we get a point cloud, where a confidence ellipse is used to encircle each observation. With this display the analyst can explore, for example, outliers that may require further investigation.

Aside from our three main novel visualizations, we  include a selection of standard diagnostic plots, such as trace, residual, and overall model fit plots, that will quickly assess aspects such as convergence and model behavior. Each of our plots can be run on any of the aforementioned \texttt{R}-packages, despite their differing formats and function arguments. 

While we make what we believe to be good default choices for the plots we produce, we provide the option to adjust many of the settings. Each aspect of the design of our plots is given careful consideration; we focus on efficient layouts, which includes both clustering and filtering, color choice, and effectively displaying uncertainty. Our new displays are appropriate for regression and classification fits and are designed to work with the three aforementioned BART packages but could readily be extended to incorporate other BART packages. Our implementation is available as the R package \texttt{bartMan} (BART Model ANalysis) which is found at \url{https://github.com/AlanInglis/bartMan}.

The outline of this paper is as follows: in Section \ref{sec:bart} we describe the formulation of a BART model and provide a brief discussion on how to access variable importance and variable interactions. In Section \ref{sec:toyExample} we describe our new visualizations for assessing variable importance and variable interactions with uncertainty, tree-based analysis, outlier identification with multidimensional scaling, and a selection of enhanced model diagnostic plots on a simple example. In Section \ref{sec:suppMat} we study BART's variable importance and variable interaction methods compared to a model agnostic approach. In Section \ref{sec:caseStudy} we demonstrate our new methods on a case study. Finally, in Section \ref{sec:conclusion} we conclude by discussing potential advantages and disadvantages of our approach, as well as potential avenues for further research.

\section{Bayesian Additive Regression Trees and Variable Importance}
\label{sec:bart}

We begin with by reviewing Bayesian additive regression trees and follow with a  review of both variable importance and variable interactions in a BART model. 

\subsection{A Short Introduction to Bayesian Additive Regression Trees}
In this section we provide a brief overview of the BART model to aid the reader in understanding our later visualizations. Those looking for a more complete description should see \cite{chipman2010bart}. BART is a Bayesian non-parametric model based on an ensemble of trees that can be used for predicting continuous and multi-class responses. Unlike regression models where a linear structure is pre-specified, BART does not assume any functional form for the model, and so automatically uncovers main and interaction effects. Given a 
continuous response variable $y_{i}$ with associated predictors $\mathbf{x}_i$, the BART model, with $m$ trees is expressed as:
\begin{align}
    y_{i} = \sum_{j=1}^{m} g(\mathbf{x}_{i}, T_{j}, M_{j}) + \epsilon_{i},
\end{align}
where $\epsilon_{i} \sim \mbox{N}(0, \sigma^{2})$ and $g(\mathbf{x}_{i}, T_{j}, M_{j}) = \mu_{j \ell}$ is a function that assigns a predicted value for the observations falling into terminal node $\ell$ of tree $j$. $T_{j}$ represents the structure/topology of tree $j$ including the split variables and the values associated with the splits $M_{j} = (\mu_{j1}, \cdots, \mu_{j b_{j}})$ represent the set of predicted values at the $b_{j}$ terminal nodes of the trees. 

The tree structure $T$ is composed of binary splitting rules of the form $[x_{j} \leq c]$, where observations which satisfy the condition go to the left and the remainder to the right. The trees are updated at each iteration in a Markov chain Monte Carlo approach where each tree structure is modified by either growing, pruning, changing, or swapping nodes. Growing a tree means that a terminal node is randomly chosen and two new terminal nodes are created, while pruning  collapses a pair of terminal nodes to their parent. A splitting rule can also be changed to a different rule, or swapped for another splitting rule in the same tree. 
In the grow and change moves a new splitting rule is required and is proposed by uniformly sampling a splitting variable and a split value though the exact generation of these rules is implementation dependent.

\begin{figure}[!t]
\begin{center}
   \begin{subfigure}{0.49\linewidth} \centering
     \includegraphics[scale=0.5]{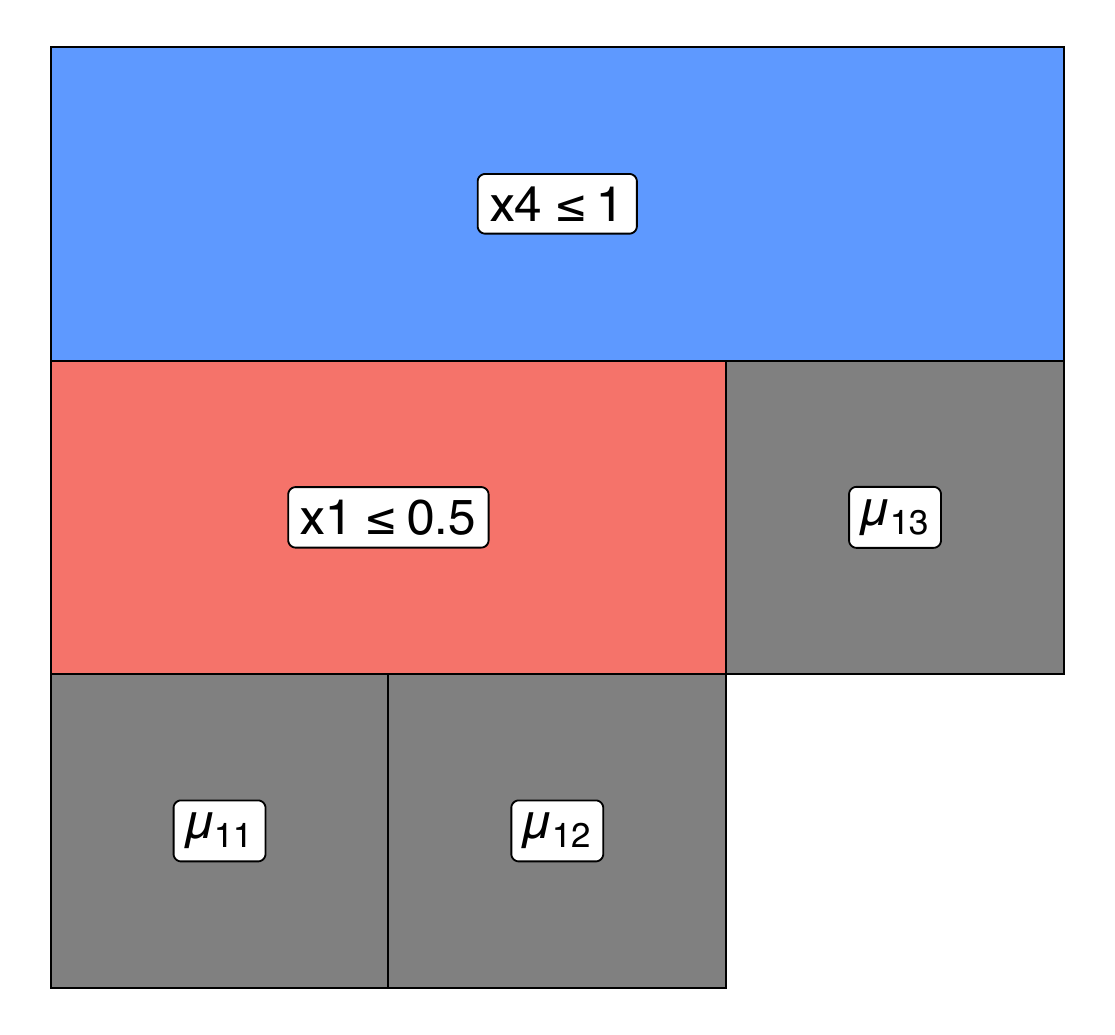} 
     \captionsetup{justification=centering}
     \caption{$T_{1}^{(1)}$}
   \end{subfigure}
   \begin{subfigure}{0.49\linewidth} \centering
     \includegraphics[scale=0.5]{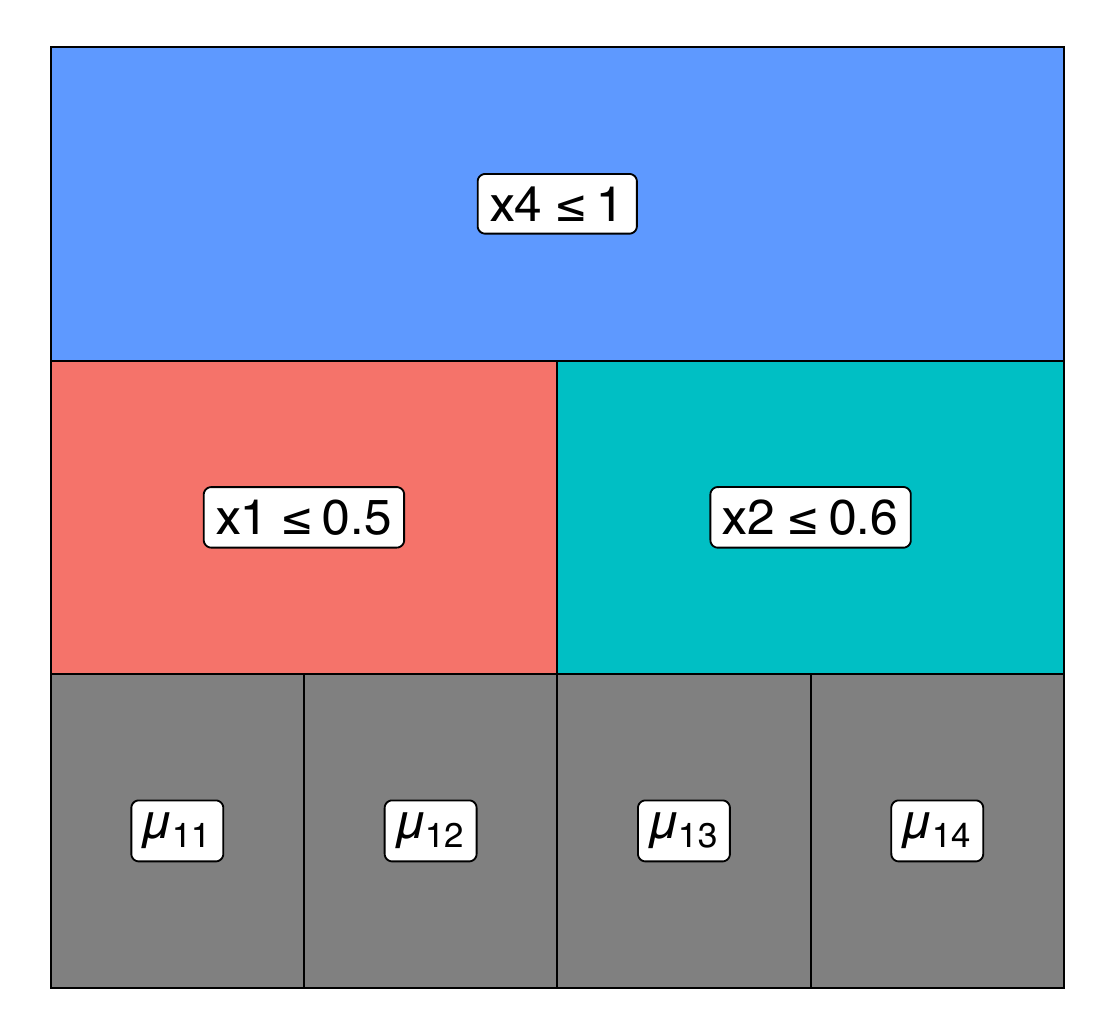} 
     \captionsetup{justification=centering}
     \caption{$T_{1}^{(2)}$}
   \end{subfigure}
    \begin{subfigure}{0.49\linewidth} \centering
     \includegraphics[scale=0.5]{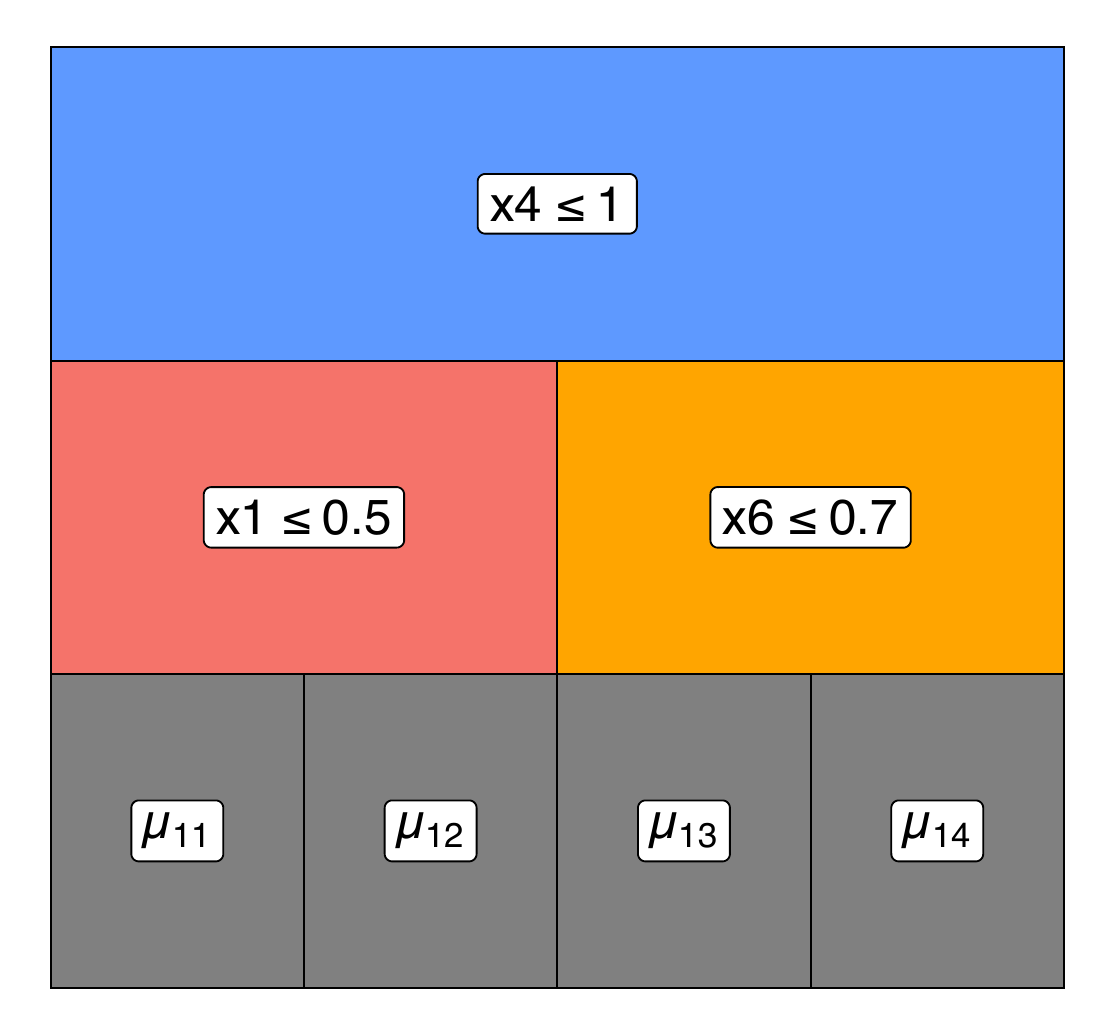} 
     \captionsetup{justification=centering}
     \caption{$T_{1}^{(3)}$}
   \end{subfigure}
    \begin{subfigure}{0.49\linewidth} \centering
     \includegraphics[scale=0.5]{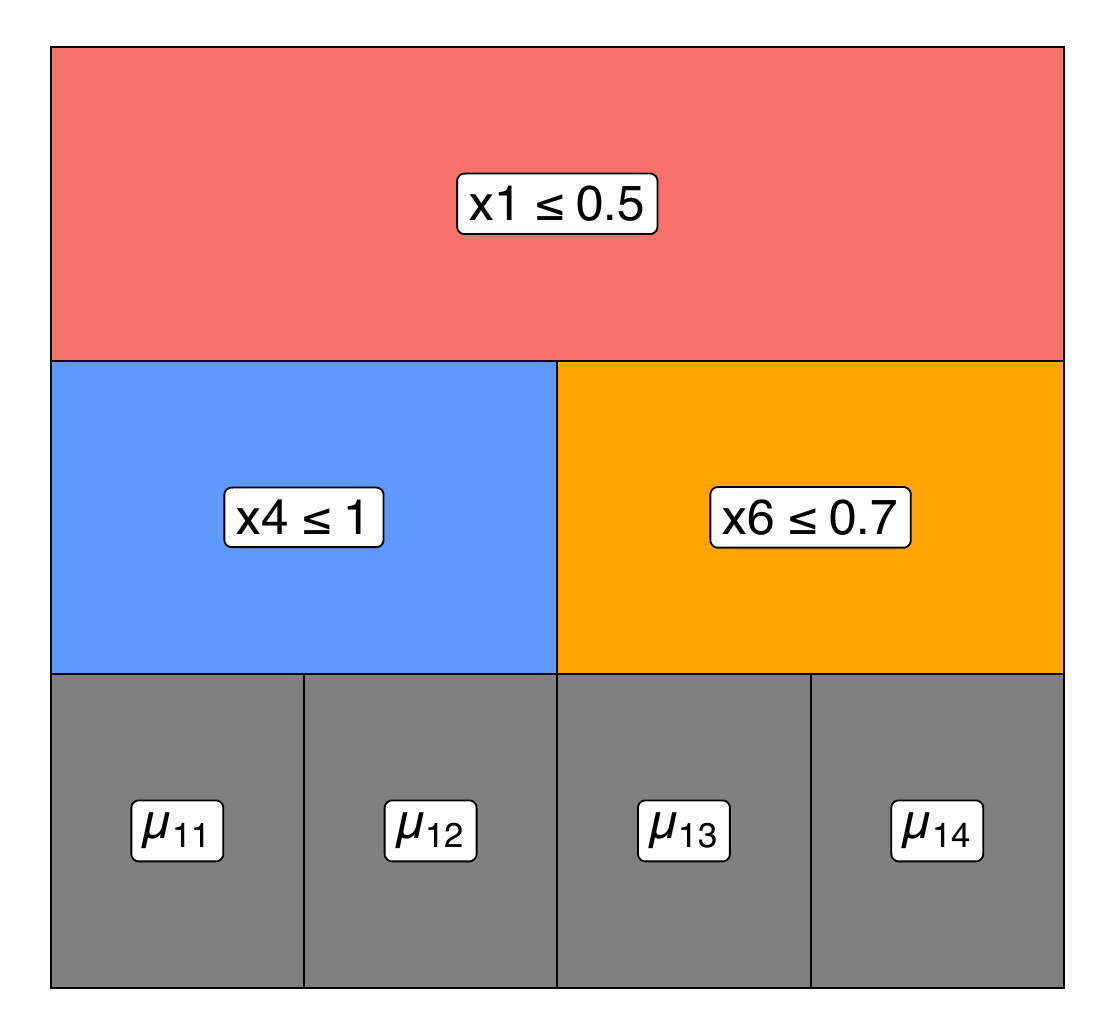} 
     \captionsetup{justification=centering}
     \caption{$T_{1}^{(4)}$}
   \end{subfigure}
\caption{An example of a tree, $T_{1}^{k}$, generated from BART over $k=1,2,3,4$ iterations. displayed as an icicle plot with the splitting rules (that is, covariates and split points) shown as colored rectangles and the terminal nodes $\mu_{j\ell}$ are shown as grey rectangles. In panel (a), $k = 1$, observations that satisfy the splitting criterion go left and tree $T_{1}^{(1)}$ has two internal nodes and three terminal nodes. Moving from panel (a) to (b) shows the grow move for the tree. Reverting from (b) back to (a) corresponds to a prune move. Panel (c) shows the change move as the splitting rule that defines $\mu_{13}$ and $\mu_{14}$ in $T_{1}^{(2)}$  is changed. Finally, in (d) the swap move can be seen when comparing the internal nodes of $T_{1}^{(3)}$  and $T_{1}^{(4)}$.} 
\label{fig:bartTREES}
\end{center}
\end{figure}

Figure \ref{fig:bartTREES} shows an example of the tree structure modifications in action. In Figure \ref{fig:bartTREES}, a tree, $T_{1}^{k}$, is generated from BART in 4 different instances, where $k=1,2,3,4$ indicates the iteration number in which the tree is updated. In the full BART model multiple trees are estimated and the predictions are created from the sum of the $\mu$ values across the trees. The tree is displayed as an icicle plot \citep{kruskal1983icicle} with the splitting rules (that is, covariates and split points) shown as colored rectangles and the terminal nodes $\mu_{j\ell}$ are shown as grey rectangles.  Icicle plots were first introduced by \cite{kruskal1983icicle} as a way to display hierarchical data in a space efficient manner. We use icicle plots to display our tree plots in later sections.

In panel (a) of Figure \ref{fig:bartTREES}
at iteration 1, observations that satisfy the splitting criterion go left and tree $T_{1}^{(1)}$ has two internal nodes and three terminal nodes. The grow move is shown going from panel (a) to panel (b), that is $T_{1}^{(1)}$ to $T_{1}^{(2)}$. An example prune move would correspond to $T_{1}^{(2)}$ reverting to $T_{1}^{(1)}$. In panel (c) we can see the change move as the splitting rule that defines $\mu_{13}$ and $\mu_{14}$ in $T_{1}^{(2)}$  is changed. Finally, in (d) the swap move can be seen when comparing the internal nodes of $T_{1}^{(3)}$  and $T_{1}^{(4)}$.

As a Bayesian model, BART adopts a set of prior distributions for the tree structure,  terminal node parameters, and residual variance. To control the depth and shape of the tree structure, a branching process prior is considered where the probability of a node being non-terminal at depth $d$ is proportional to $\alpha (1 + d)^{-\beta}$, where $\alpha \in (0,1)$ and $\beta \geq 0$. \cite{chipman2010bart} recommend $\alpha = 2$ and $\beta = 0.95$ as default, which favors shallow and balanced trees. A side effect of this choice is that noise (i.e. uninformative) variables are often chosen in shallower tree structures as the tree prior can outweigh the likelihood when a large number of trees is used. The terminal node parameters $\mu_{j \ell}$ are assumed to be independent and identically distributed, that is,  $\mu_{j \ell} \sim \mbox{N}(0, \sigma^{2}_{\mu})$, where $\sigma^{2}_{\mu}$ is the residual variance of the terminal node parameters, which is usually fixed. The value of $\sigma^{2}_{\mu}$ is usually set with the aim of forcing the trees to be shallow and shrink their predictions towards zero so that each tree only contributes a small amount to the overall prediction. Finally, the prior on the residual variance $\sigma^{2}$ is an Inverse Gamma.

Posterior sampling is based on a Metropolis-within-Gibbs MCMC structure where the trees are sequentially updated through partial residuals. For one MCMC iteration, each tree in the ensemble is modified  and then compared to its previous version via a Metropolis-Hastings update. The update  involves a marginalized likelihood and the tree prior. The marginalized likelihood is an essential element to avoid trans-dimensional MCMC, and simplifies computation. Given the tree structure, all terminal node parameters $\mu_{j \ell}$ are updated based on a closed-form posterior conditional distribution. After updating all trees, the variance $\sigma^{2}$ is updated; a more complete description can be found in \cite{chipman2010bart} and \cite{ tan2019bayesian}.

In the above, we have described the BART model for a univariate and continuous response variable. However BART has been extended into many different areas, such as survival analysis \citep{sparapani2016nonparametric, linero2021bayesian}, time series analysis \citep{starling2020bart}, multivariate skewed response \citep{um2021bayesian}, and high-dimensional data \citep{hernandez2018bayesian, lineroAnDyang2018, he2018xbart}. While in this work we do not apply our methods to these extensions of BART, there is in principle no reason why our methodology could not be extended to incorporate the above BART extensions.

\subsection{Variable Importance and Variable Interaction Calculations with BART}
\label{sec:viviBARTintro}
Variable importance is a measure of a single variable's impact on the response. Multiple methods exist for evaluating variable importance, depending on the model; for a comprehensive review of different variable importance techniques see \cite{wei2015variable}. \cite{chipman2010bart} propose a method called the inclusion proportion to evaluate the variable importance in a BART model from the  posterior samples of the tree structures. Their measure of variable importance first calculates for each iteration the proportion of times a variable is used to split nodes considering all $m$ trees, and then averages these proportions across all iterations.

More formally, let $K$ be the number of posterior samples obtained from a BART model. Let $c_{rk}$ be the number of splitting rules using the $r$th predictor as a split variable in the $k$th posterior sample of the trees' structure across $m$ trees. Additionally, let $c._k = \sum_{r = 1}^{p} c_{rk}$ represent the total number of splitting rules found in the $k$th posterior sample across the total $p$ variables. Therefore, $z_{rk} = c_{rk}/c._k$ is the proportion of splitting rules for the $r$th variable, and the average use per splitting rule is given by:
\begin{align}
     \mbox{VImp}_{r} = \frac{1}{K} \sum_{k=1}^{K} z_{rk}
     \label{eqn:vip}
\end{align}
However \cite{chipman2010bart} noted that this method of evaluating importance is less effective when the number of trees, $m$, is large because weakly influential predictor variables can be added to the tree structure and so may provide spurious importance values for the non-important variables. As $m$ decreases this effect is diminished because the less important variables get swapped out of the trees for more informative variables. 

Variable interaction is generally considered as when a pair (or more) of variables jointly impact on the response. In our work we focus on bivariate interactions only. \cite{BARTMachPack} suggested a measure of interaction obtained by observing successive splitting rules in each tree. 
Let $c_{rqk}$ be the number of  splitting rules  using predictors $r$ and $q$ successively (in either order) in the $k$th posterior sample. Additionally, let $c.._k = \sum_{r = 1}^{p} \sum_{q = 1}^{p} c_{rqk}$ represent the total number of successive splitting rules found in the $k$th posterior sample. We follow the convention of \cite{BARTMachPack} and we treat the order of successive splits as not important and we sum the $r,q$ counts with the $q,r$ counts. Therefore, the proportion $z_{rqk} = c_{rqk}/c.._k$, provides an estimate of the  interaction between variables $r$ and $q$:
\begin{align}
    \mbox{VInt}_{rq} = \frac{1}{K} \sum_{k=1}^{K} z_{rqk}.
    \label{eqn:vint}
\end{align}

As this method follows a similar technique to evaluating the inclusion proportion, the same pitfalls noted by  \cite{chipman2010bart} apply,  namely that the prior distribution may favor trees containing successive  predictor variables where there is no true interaction present if the number of trees is large. For a comparison of both the variable importance and variable interaction methods against a model agnostic approach for evaluating these metrics, see Section \ref{sec:suppMat}.

It should be noted that if any of the variables used to build the BART model are categorical, the aforementioned BART packages replace the categorical variables with $d$ dummy variables, where $d$ is the number of factor levels. For some of our plots, the inclusion proportions for variable importance and interaction are then adjusted by aggregating over factor levels. This provides a complete picture of the importance of a factor, rather than that associated with individual factor levels.

Since both the VImp and VInt values are calculated from the full posterior, it is trivial to compute an uncertainty associated with their measurement, simply by storing the importance and interaction calculations per iteration. These can be summarized by the usual means by which posterior distributions are analyzed. We will use uncertainty metrics obtained from these distributions in our variable importance and interaction displays of Section \ref{sec:toyExample}. 

\section{New visualizations for BART}
\label{sec:toyExample}
To illustrate our new visualizations we use a subset of the iris data \citep{iris} where the response is binary and made up of two species (that is, setosa and versicolor). We then fit a BART model to the data using \texttt{bartMachine}, using the default setting of 1000 iterations with a burn-in of 250. For simplicity of exposition we set the number of trees to be 20.

We introduce the following visualizations: improved plots of variable importance and interaction which include the uncertainty induced by the posterior distribution of trees; plots of the tree structures which show the splitting variables, the split distribution, and the terminal node values; the ability to identify outlying and influential observations through the terminal node proximity matrix and multi-dimensional scaling; and a set of enhanced model diagnostics for identifying convergence and performance issues. 

\subsection{Variable Importance and Interaction with Uncertainty}
\label{sec:viviBart}

\begin{figure}[!b]
\centering
\includegraphics[scale = 0.5]{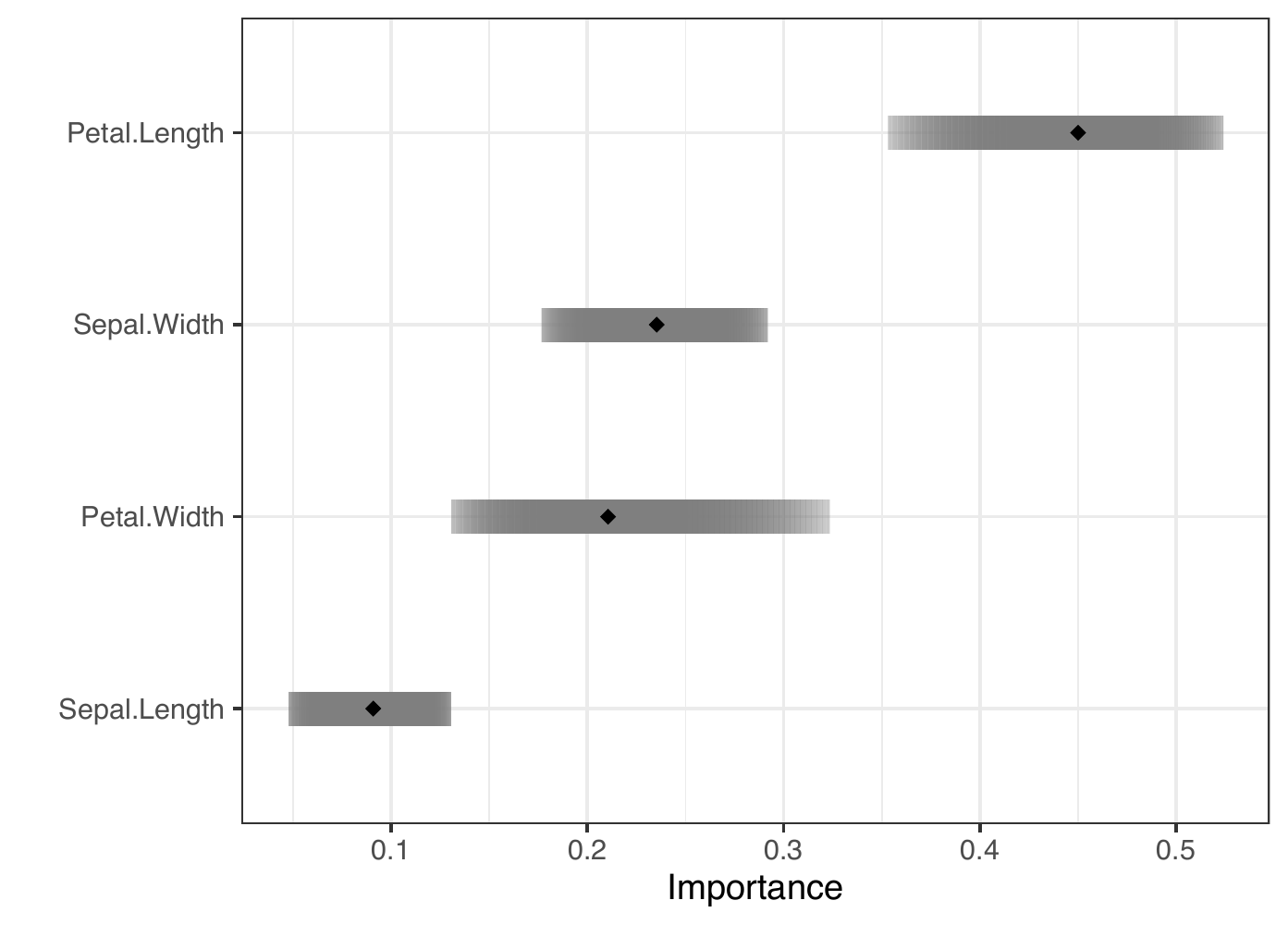}
\caption{Inclusion proportions for the iris data are shown with the 25\% to 75\% quantile interval extending from the points. Here Petal.Length is ranked as the most important variable.}
\label{fig:vimpIris}
\end{figure}

In this section we present visualizations of the variable importance methods described in Section \ref{sec:viviBARTintro}. In Figure  \ref{fig:vimpIris} 
we show the median of the inclusion proportion as a black point, with the variables ordered from the largest median importance measure (at the top) and descending. In this case the 25\% to 75\% quantile interval extending from each point is displayed as a grey bar. We can see that Petal.Length is the most important variable and that Sepal.Width and Petal.Width have similar inclusion proportions. Sepal.Width importance has a lower  degree of uncertainty, as indicated by the relatively small quantile interval, whereas the  Petal.Width importance has a large quantile interval associated with it, and therefore its importance measure should be viewed with a level of caution.

In \cite{inglis2022visualizing}, the authors propose using a heatmap to display both importance and interactions simultaneously, where the importance values are on the diagonal and interaction values on the off-diagonal. The advantage of such a display is that it allows one to easily identify which variables are relevant as separate predictors while also seeing which variable pairs have high interaction. This method, coupled with the seriation technique described by \cite{inglis2022visualizing}, brings predictors with high importance and interaction to the top-left of the heatmap and less relevant predictors to the bottom-right.  

Here we adapt the heatmap displays of  importance and interactions  to include the uncertainty using a VSUP.
The colors for the VSUP heatmap were carefully chosen to be distinguishable, color-blind friendly, and to aid in highlighting high values, while still making the uncertainty prominent. To achieve this, we follow the advice of \cite{strode2019operationalizing}, who build upon the work of \cite{trumbo1981theory}, and aim to highlight and focus the reader's attention on the interesting data. 

\begin{figure}[!b]
\begin{center}
   \begin{subfigure}[t]{0.49\linewidth} 
   \centering
     \includegraphics[scale=0.34]{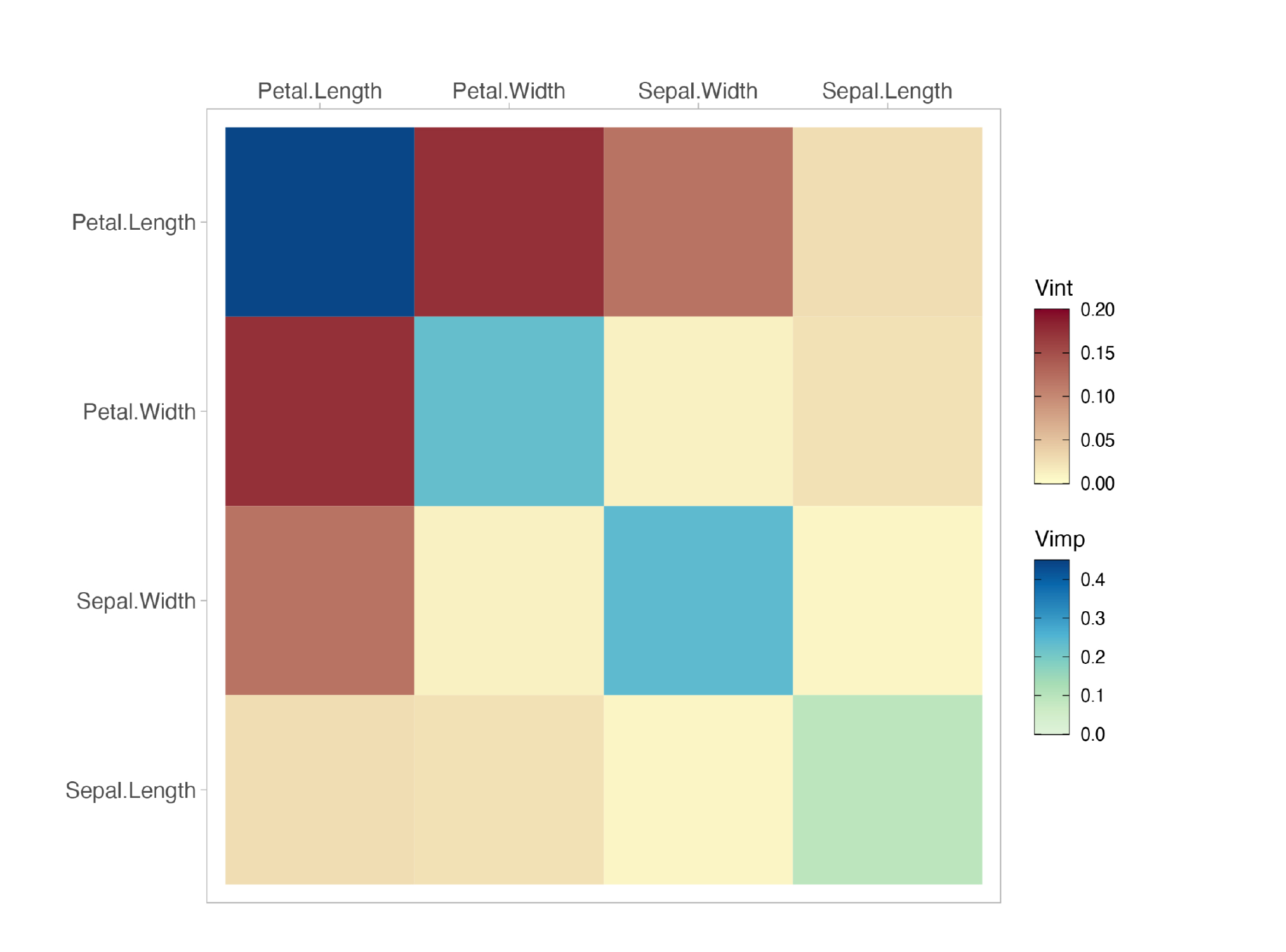} 
     \captionsetup{justification=centering}
     \caption{}
   \end{subfigure}
   \begin{subfigure}[t]{0.49\linewidth} 
   \centering
     \includegraphics[scale=0.34]{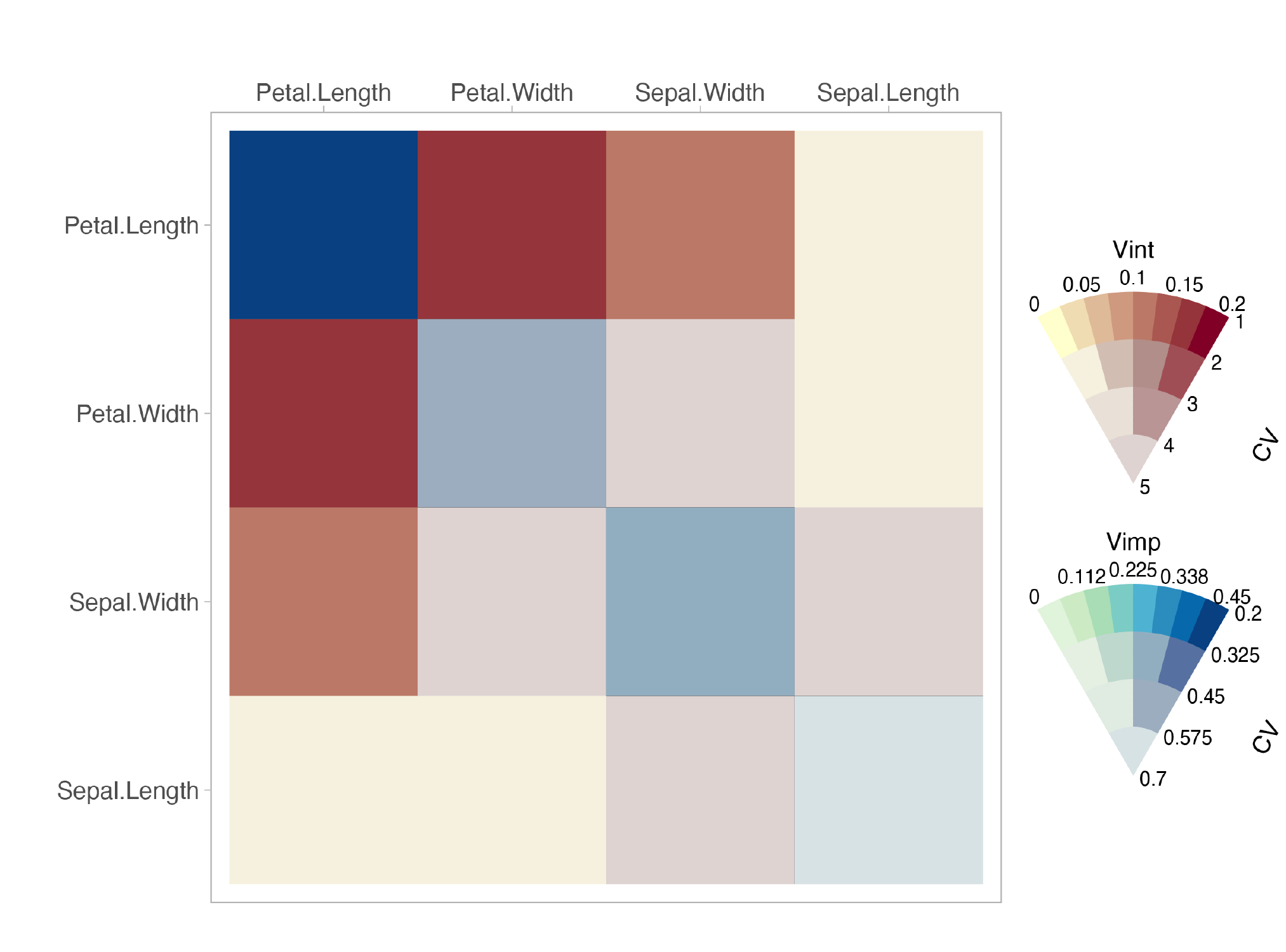} 
     \captionsetup{justification=centering}
     \caption{}
   \end{subfigure}
\caption{In (a) the importance values are on the diagonal and interaction values on the off-diagonal. Petal.Length is the most important variable and there is a strong interaction between Petal.Length and Petal.Width. In (b) the same values values are shown but with the coefficient of variation included by use of a VSUP. Both the importance measure of Petal.Length, and the interaction measure between Petal.Length and Petal.Width have  low coefficient of variation.
}
\label{fig:vivi}
\end{center}
\end{figure}

Figure \ref{fig:vivi} 
presents a comparison of  heatmaps showing the importance and interactions jointly with and without uncertainty.
In both heatmaps the variable importance is displayed on the diagonal and the interactions on the off-diagonal. In (a), we can see that Petal.Length is the most important variable when predicting Species. There also appears to be a strong interaction between Petal.Length and Petal.Width. In (b) the same values are shown but with a measure of uncertainty included, in this case the coefficient of variation (CV). Other error metrics such as standard deviation can be applied, though in the case of using proportions larger values tend to have greater uncertainty and so our preference is for the CV. In both (a) and (b) the same method is used to obtain the importance and interaction scores, resulting in comparable scales. Comparing the two plots we observe that in (b) the most important variable, Petal.Length, has a small variation relative to its mean. The Petal.Length and Petal.Width interaction value has a low coefficient of variation and is consequently highlighted in (b), whereas  Petal.Length and Sepal.Length have a low interaction score with relatively high variation.

\subsection{Tree-Based Plots}
\label{sec:treeToy}
In this section we examine more closely the structure of the decision trees created when building a BART model. Examining the tree structure may yield information on the stability and variability of the tree structures as the algorithm iterates to create the posterior. By sorting and coloring the trees appropriately we can identify important variables and common interactions between variables for a given iteration. Alternatively we can look at how a single tree evolves through the iteration to explore the fitting algorithm's stability. 

\begin{figure}[!b]
\centering
\includegraphics[scale = 0.57]{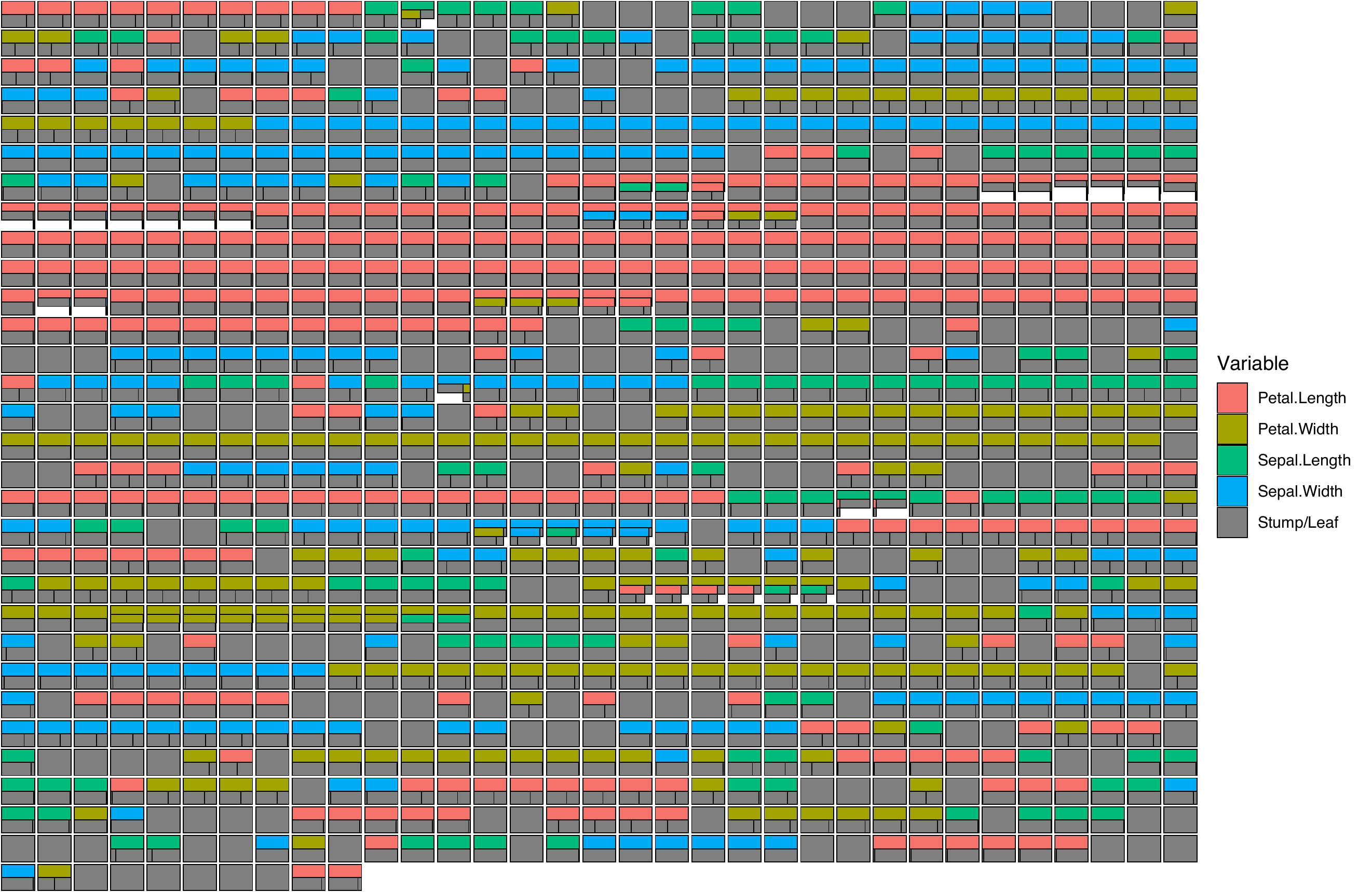}
\caption{A single tree over 1000 iterations. The colored bars indicate which variable is used for the split at that point. Grey boxes indicate stumps or terminal nodes. The vertical black lines in the terminal nodes indicate the proportion of the data being split into the left or right terminal node. }
\label{fig:allTrees}
\end{figure}

In Figure \ref{fig:allTrees}, 
we show how a single selected tree changes over all 1000 post burn-in iterations. We use an icicle plot to display the trees. As noted by \cite{barlow2001comparison}, icicle plots are preferred by users when compared to other methods to display decision trees and use space more efficiently. Additionally, the number of observations within each decision tree node is represented in icicle plots by scaling the node size accordingly. In Figure \ref{fig:allTrees}, each parent node is colored according to the variable with the terminal nodes all colored a dark grey. A stump is represented by a solid grey square (although stumps can be removed from the plots if desired). (More options to color the nodes by certain parameters are shown in later plots in this section.)  With this display we see  how a tree evolves over iterations. Here we see the prevalence of Petal.Length as a splitting variable (red rectangles) once again indicating the importance of this predictor. Additionally, most iterations have a single split on the root node, with very few trees showing an interaction. As the nodes are sized according to the number of observations, we observe that in the seventh and eighth rows some trees have an empty, white space. In this case most of the observations fall into the single terminal node on the left. The remaining observations go right and split again.

\begin{figure}[!t]
\begin{center}
   \begin{subfigure}{0.49\linewidth} \centering
     \includegraphics[scale=0.32]{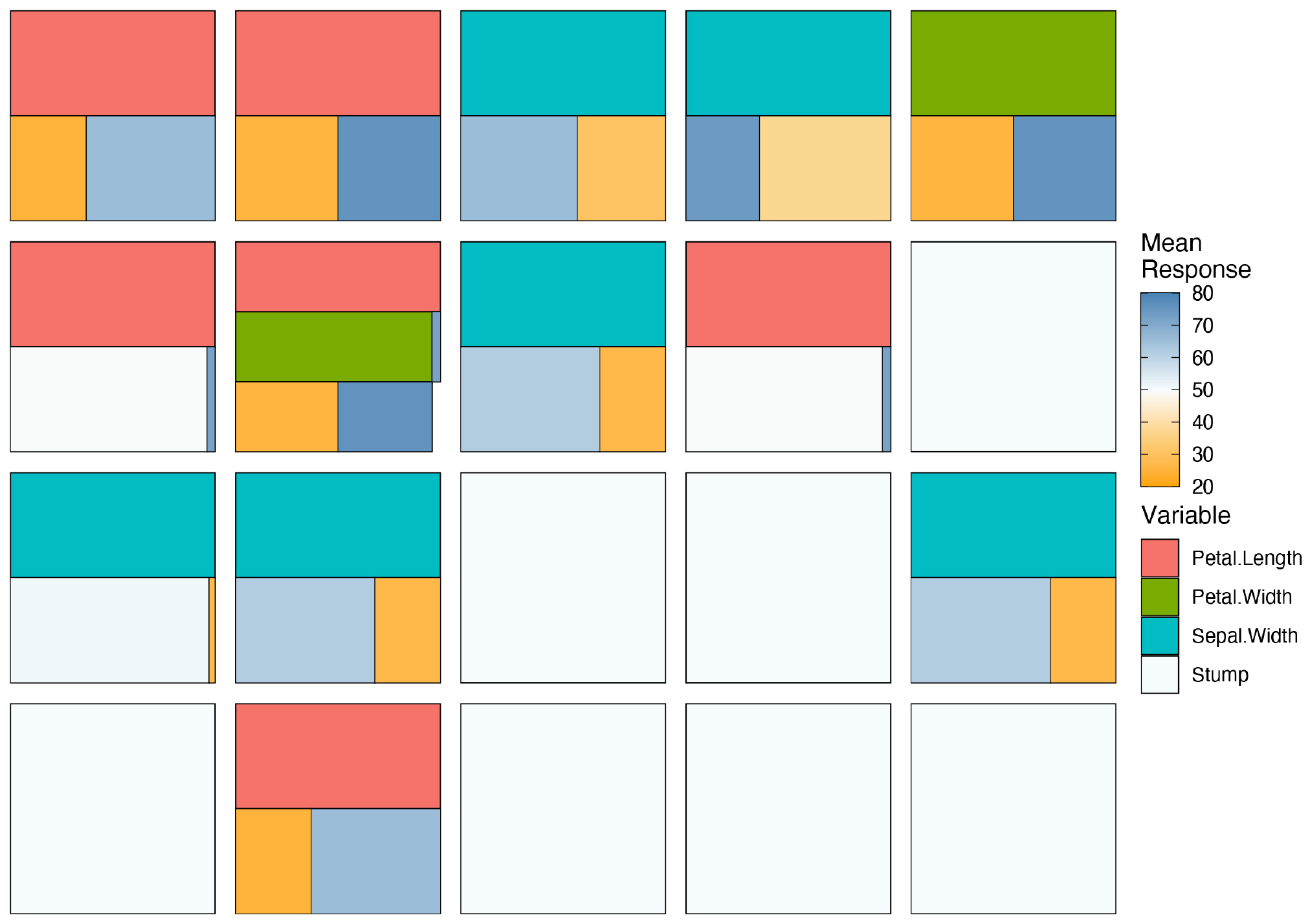} 
     \captionsetup{justification=centering}
     \caption{}
     \label{fig:treeA}
   \end{subfigure}
   \begin{subfigure}{0.49\linewidth} \centering
     \includegraphics[scale=0.32]{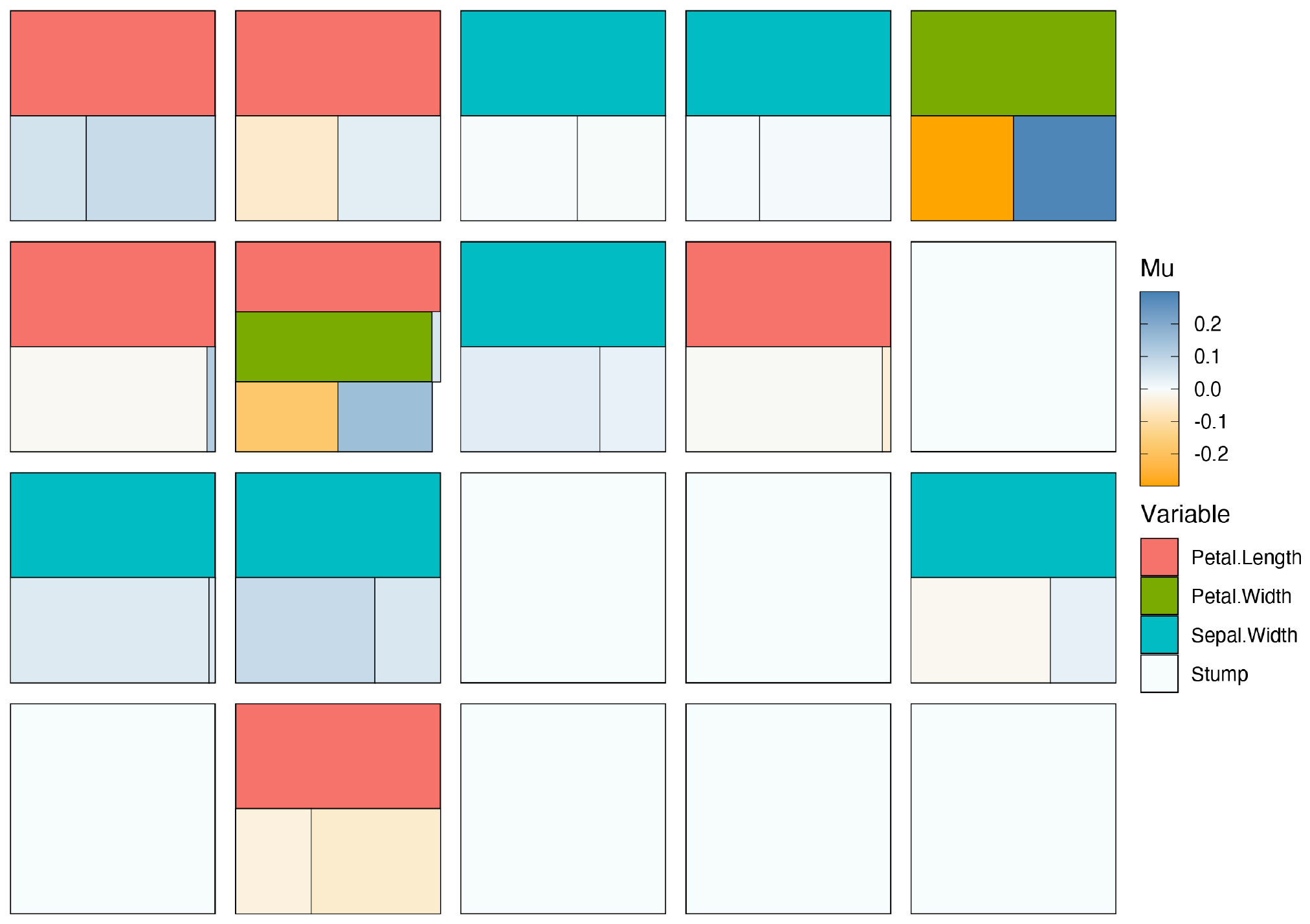} 
     \captionsetup{justification=centering}
     \caption{}\label{fig:treeB}
   \end{subfigure}
    \begin{subfigure}{0.49\linewidth} \centering
     \includegraphics[scale=0.32]{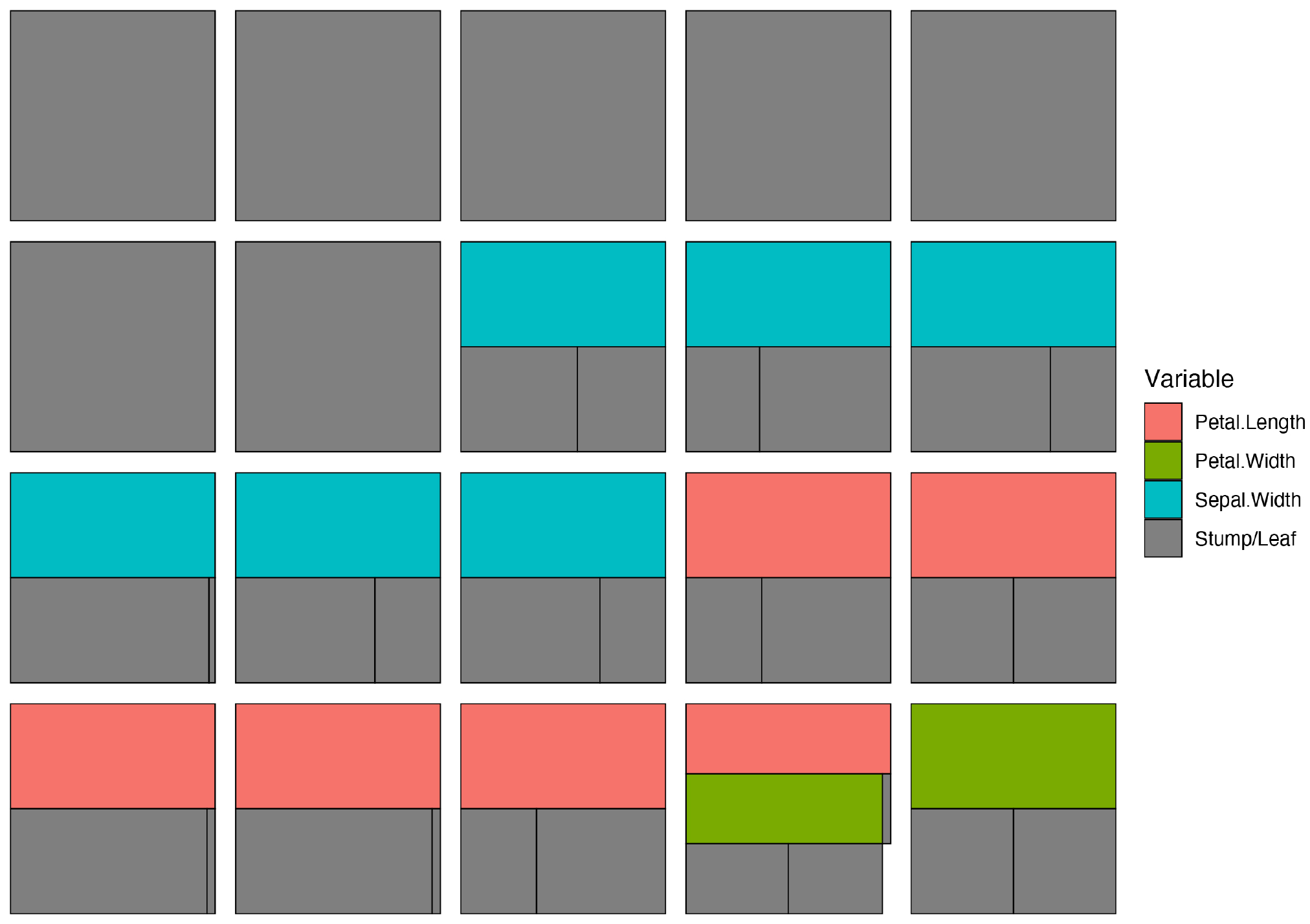} 
     \captionsetup{justification=centering}
     \caption{}\label{fig:treeC}
   \end{subfigure}
    \begin{subfigure}{0.49\linewidth} \centering
     \includegraphics[scale=0.32]{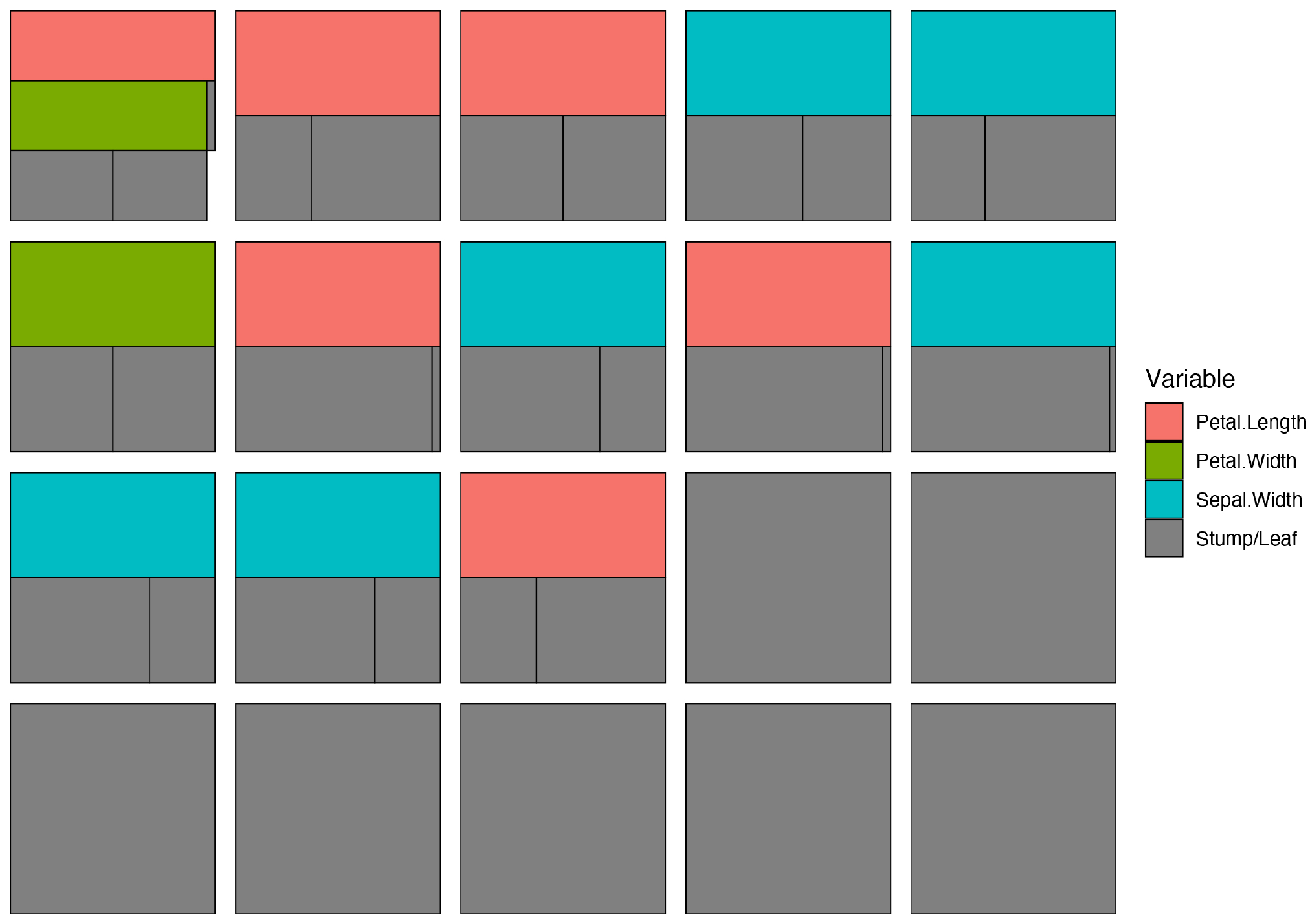} 
     \captionsetup{justification=centering}
     \caption{}\label{fig:treeD}
   \end{subfigure}
\caption{All trees in a selected iteration. In (a) the terminal nodes and stumps are colored by the mean response. In (b) the terminal nodes and stumps are colored by the predicted value $\mu$. In (c) we sort the trees by structure starting with the
most common tree and descending to the least common tree shape and in (d) we sort the trees by tree depth.} 
\label{fig:treesExample}
\end{center}
\end{figure}

In our tree displays, it is also useful to view different aspects or metrics. In Figure \ref{fig:treesExample} 
we explore some of these aspects by displaying all the trees in a selected iteration (in this case, we chose the iteration with lowest residual standard deviation). We consider variations which color terminal nodes and stumps by the mean response (panel (a)), color them by the terminal node parameter value (panel (b)), sort the trees by structure starting with the most common tree and descending to the least common tree found for easy identification of the most important splits (panel (c)), or sort the trees by depth (panel (d)). As the $\mu$ values in (b) are centered around zero, we use a single-hue, colorblind friendly, diverging color palette to display the values. For comparison, we use the same palette to represent the mean response values in (a).

Different interesting findings are seen in the four panels. Panel (b) indicates that tree 5 (the top right tree displayed) has a much greater influence on the overall predictions than the others, which seems surprising given the nature of the shrinkage prior used in BART which aims to shrink the terminal node parameters towards zero. From (c) we  observe that the most common tree structure in this iteration is actually a stump. The most common non-stump tree type has Sepal.Width as the root with a single binary split. Furthermore, in this iteration Petal.Length and Sepal.Width are both used as a splitting variable an equal number of times. In (d) it is quickly identified that the vast majority of trees in this iteration have one or zero splits. 

When the number of variables or trees is large it can become harder to identify interesting features. We provide a plot that can be used to highlight interesting features by accentuating selected variables by coloring them brightly while uniformly coloring the remaining variables a light grey. When coupled with the sorting shown previously in Figure \ref{fig:treesExample} we have found that this more clearly identifies relationships of interest. As the iris data has very few predictors, we omit this plot here but an example of it can be seen the larger case study example of Figure \ref{fig:bs_sel} in Section \ref{sec:caseStudy}.

\begin{figure}[!t]
\centering
\includegraphics[scale = 0.4]{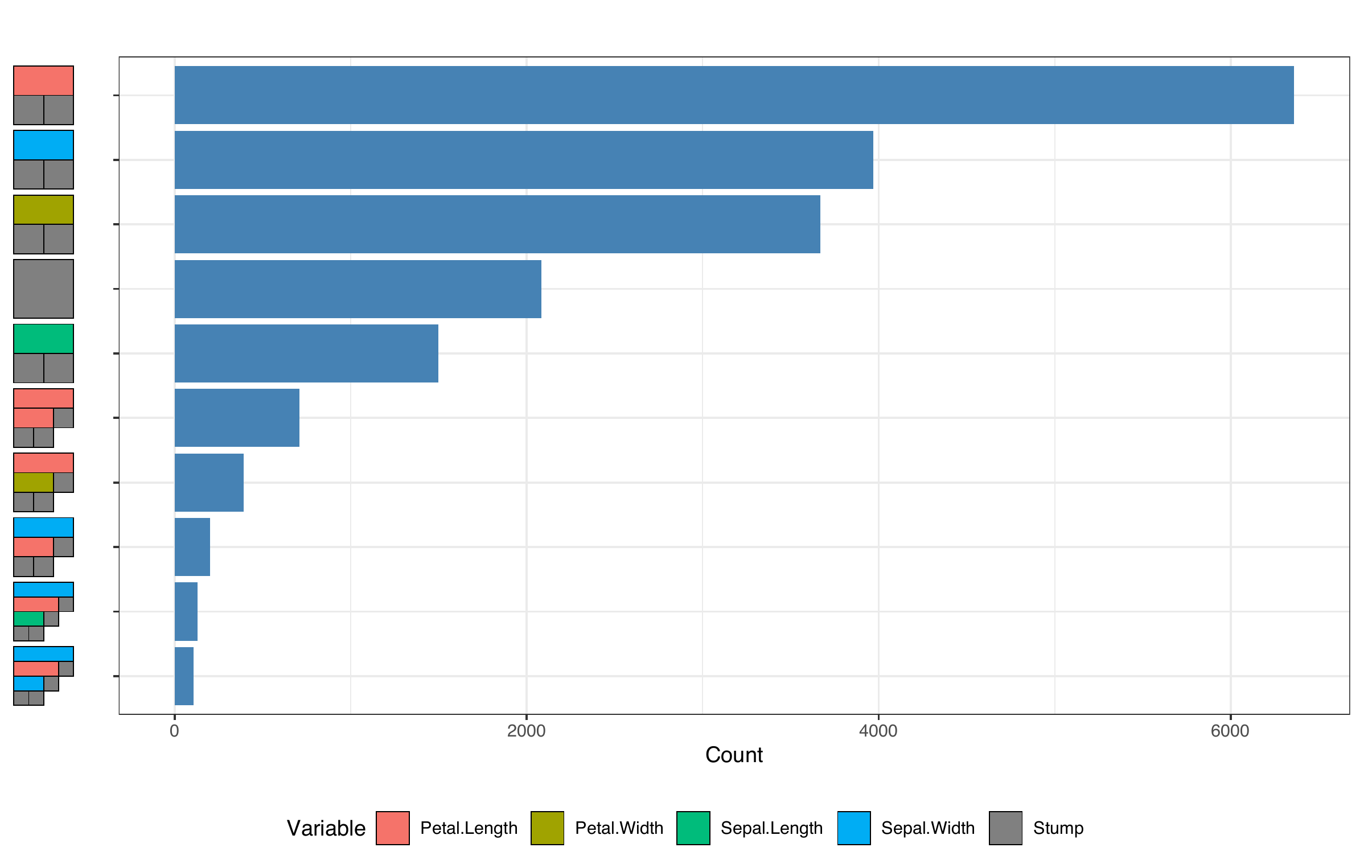}
\caption{Bar plot of the top 10 most frequent tree types over all iterations. Trees with a single binary split on Petal.Length occur the most often.}
\label{fig:irisBarplot1}
\end{figure}

Finally, as an alternative to the sorting of the tree structures, seen in Figure \ref{fig:treesExample} (c), we provide a bar plot summarizing the tree structures. Figure \ref{fig:irisBarplot1} 
shows a barplot of the frequency of the tree types over all iterations, filtered to show the top 10 most frequent trees, where the legend indicates the tree structure with the node sizes equally proportioned. To count the tree structures, we use the same sorting algorithm as Figure \ref{fig:treesExample} (c). This seems most useful when summarizing a large number of trees (though again these plots can also be created for a single tree across iterations or to display all trees in a single iteration).  We can see that the most common tree type over all iterations is the tree that has a single binary split on Petal.Length, with the second most common being the tree that has a single binary split on Sepal.Width. Additionally, we can see that Petal.Length appears in several of the other top 10 most common tree structures. This is in agreement with the inclusion proportion variable importance plot of Figure \ref{fig:vimpIris} which tells us that Petal.Length is used as a splitting rule most often.

\subsection{Outlier Identification with Multidimensional Scaling}
\label{sec:mds}

Proximity matrices combined with multidimensional scaling (MDS) are commonly used in random forests to identify outlying observations \citep{breiman2001randomforest}. Both proximites and MDS have been shown to be useful tools and can be applied to wide range of data types, including genomic and ecological data \citep[for example, see ][]{englund2012novel,cutler2007random}. However, to our knowledge, these methods have not yet been  implemented for a BART model. When two observations lie in the same terminal node repeatedly they can be said to be similar, and so an $N\times N$ proximity matrix is obtained by accumulating the number of times at which this occurs for each pair of observations, and subsequently divided by the total number of trees. A higher value indicates that two observations are more similar. The proximity matrix is then visualized using classical MDS (henceforth MDS) to plot their relationship in a lower dimensional projection. 

\begin{figure}[!t]
\centering
\includegraphics[scale = 0.4]{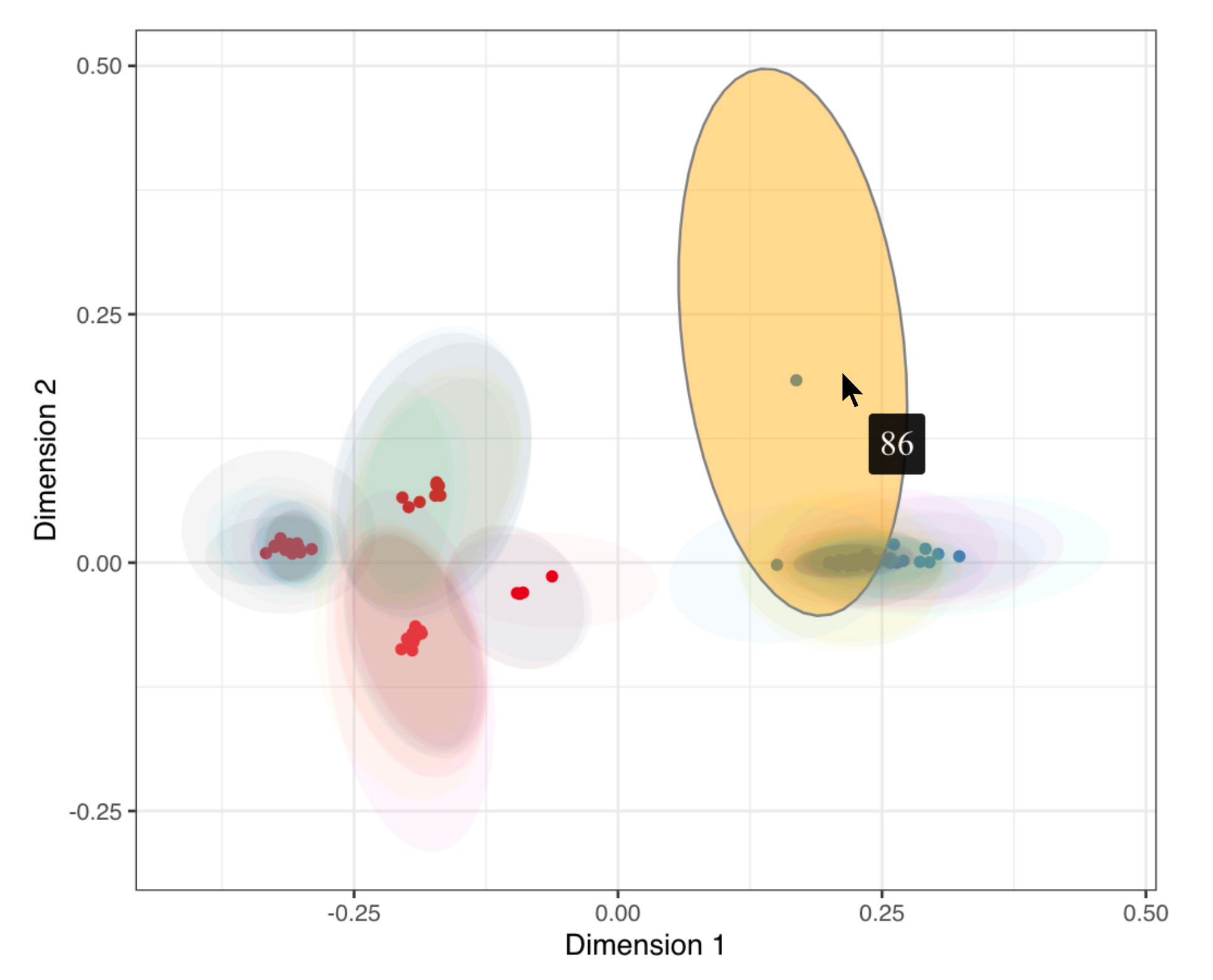}
\caption{Interactive MDS plot of the iris data where the points are colored by class (in this case, either Species). Each 95\% confidence ellipse corresponds to each observation's posterior location. When hovering the mouse pointer over an ellipse, the ellipse is highlighted and the observation is displayed.}
\label{fig:mdsInter}
\end{figure}

In BART there is a proximity matrix for every iteration and thus a posterior distribution of proximity matrices. While trivial to then apply MDS to each matrix we introduce a rotational constraint so that we can similarly obtain a posterior distribution of each observation in the lower dimensional space. We first choose a target iteration (we use the iteration with lowest residual standard deviation) and apply MDS. For each subsequent iteration we rotate the MDS solution matrix to match this target as closely as possible using Procrustes' method. We end up with a point for each observation per iteration per MDS dimension. We then group the observations by the mean of each group and  produce a scatterplot, where each point represents the centroid of the location of each observation across all the MDS solutions. This allows for an easier to read estimate of potentially outlying data points. We extend this further by displaying the 95\% confidence ellipses around each observation's posterior location in the reduced space.  Since these are often overlapping  we have created an interactive version that highlights an observation's ellipse when hovering the mouse pointer above the ellipse (Figure \ref{fig:mdsInter} shows a screenshot of this interaction in use). The observation number is also displayed during this action.

 In Figure \ref{fig:mdsInter},
each point represents the centroid of the location of each observation across all the MDS solutions and are colored according to their class (in this case, either Species). We can see that most of the variability is, unsurprisingly, in the first-dimension, and while some points have quite different posterior distributions, the uncertainty on many of them is large. Observation 86 appears to have a large uncertainty and is separated by some distance from the other observations in that class. This is an interesting finding as previous outlier detection studies using the iris data (such as \cite{acuna2004meta} and \cite{liu2015ensemble}) have not identified this observation as an outlier.
 Further investigation, by examining the tree structure and a proximity matrix plot (which, in the interest of space, we omit here) show that this observation is commonly found in the same nodes as those observations from the other class.

\subsection{Enhanced BART model diagnostics}
\label{sec:diagPlots}
In this section, we examine some of the more common issues a researcher may face when running a BART model. These include checking for convergence, the stability of the trees, the efficiency of the algorithm, and the predictive performance of the model. In our experience, most popular BART R packages are limited in scope for creating informative model visualizations (with the possible exception of \texttt{bartMachine} which features versions of Figures \ref{fig:toyAccept} and \ref{fig:depthAndNodes}). Our goal in these plots is to provide a convenient and useful summary of the model's characteristics which is invariant to the choice of package. A useful side effect of these plots is the ability to compare BART fits from different BART R packages. In the following section we show a selection of diagnostic plots using both the \texttt{bartMachine} and \texttt{dbarts} packages to build our models. Both models have the same hyperparameters of 1000 iterations with a burn-in of 250 and 20 trees. We use the same two-species subset of the iris data as before.

\subsubsection{Acceptance Rate of Trees}
As discussed in Section \ref{sec:bart}, BART uses a Metropolis-Hastings algorithm to determine the type of tree structure accepted at each tree in each MCMC iteration.  The trees are individually modified by either a grow, prune, change, or swap step and compared to its previous version by calculating the acceptance ratio. The acceptance rate is therefore measured as the percentage of accepted proposed trees across the iterations. 

\begin{figure}[!b]
\begin{center}
   \begin{subfigure}{0.49\linewidth} 
   \centering
     \includegraphics[scale=0.37]{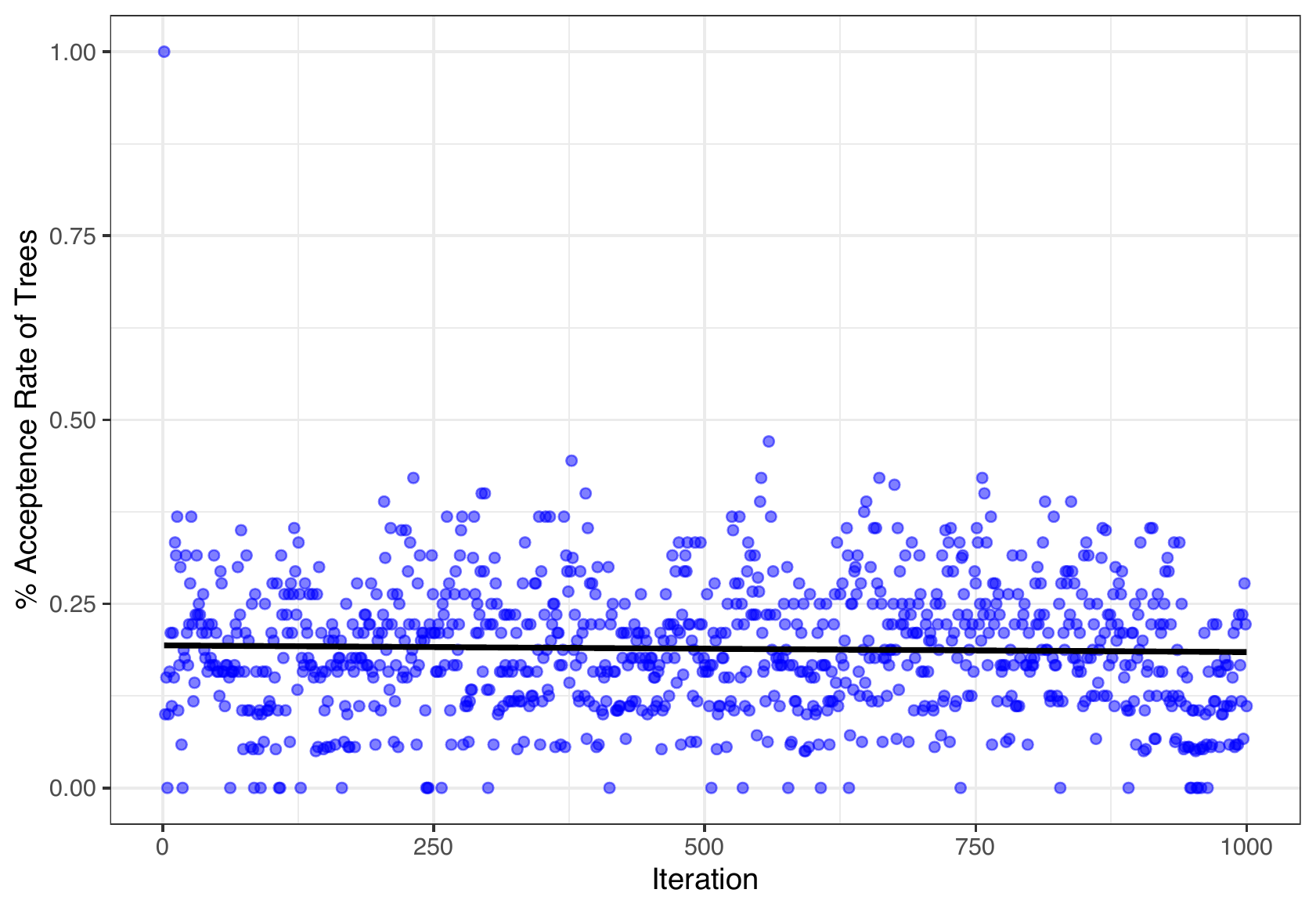} 
     \captionsetup{justification=centering}
     \caption{Acceptance rate per iteration using \texttt{bartMachine}}
   \end{subfigure}
   \begin{subfigure}{0.49\linewidth} 
   \centering
     \includegraphics[scale=0.37]{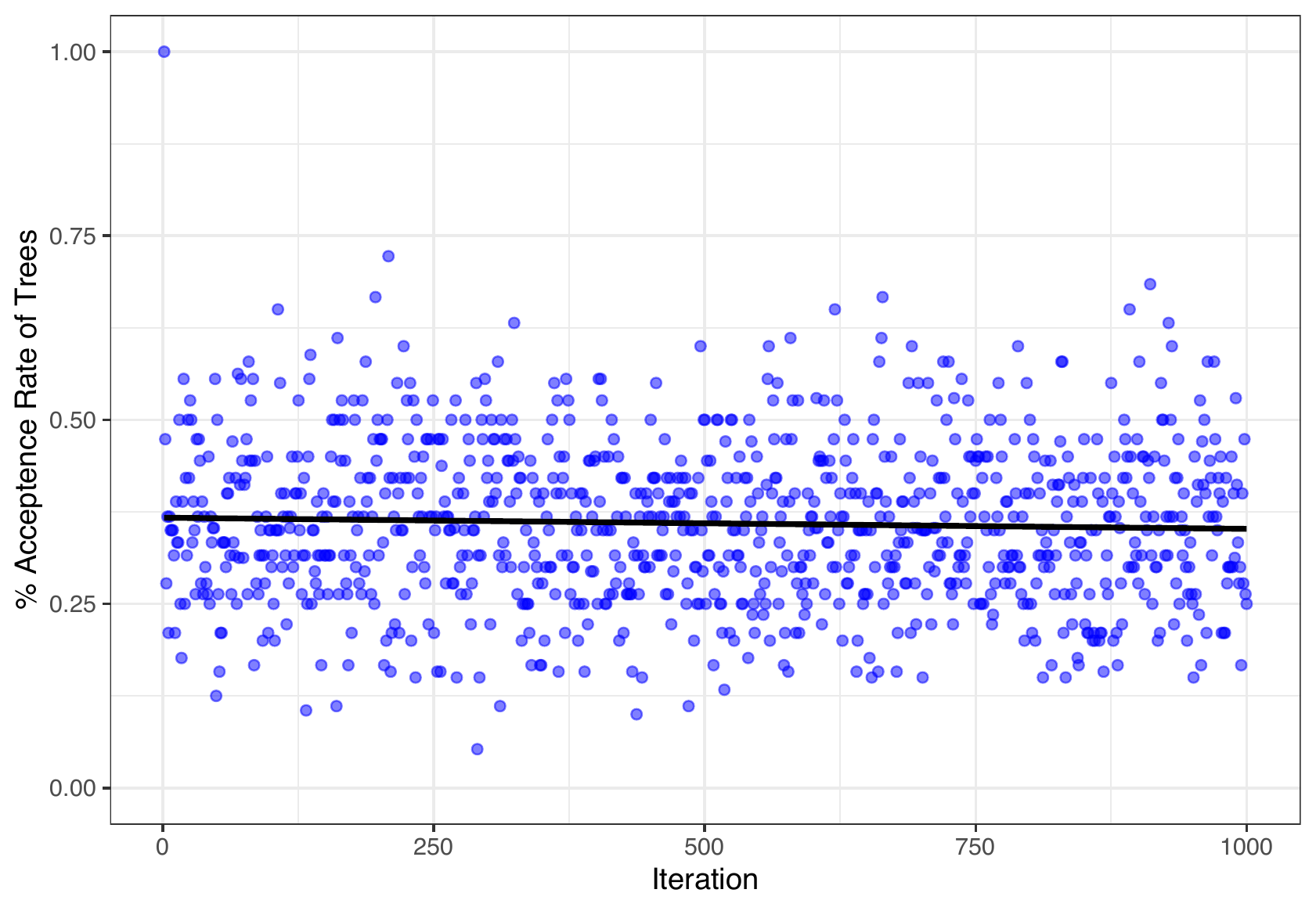} 
     \captionsetup{justification=centering}
     \caption{Acceptance rate per iteration using\\ \texttt{dbarts}}
   \end{subfigure}
\caption{Post burn-in acceptance rate of trees per iteration for a \texttt{bartMachine} and \texttt{dbarts} fit in (a) and (b), respectively. A black regression line is shown to indicate the changes in acceptance rate across iterations and to identify the mean rate. We can see that the \texttt{dbarts} fit has a higher acceptance rate than the \texttt{bartMachine} fit.} 
\label{fig:toyAccept}
\end{center}
\end{figure}

Figure \ref{fig:toyAccept} 
shows the post burn-in percentage acceptance rate across 1000 iterations for both BART models, where each point represents a single iteration. A regression line is shown to indicate the changes in acceptance rate across iterations and to identify the mean rate. Both plots are forced to display the same vertical axis range. Clearly there is a higher acceptance rate (approx 35\%) in the \texttt{dbarts} fit. None of the iterations in \texttt{dbarts}  have zero trees accepted, while this occurs commonly for \texttt{bartMachine}. This can also be seen in Figure \ref{fig:allTrees} where there are runs of identical trees, indicating that no new trees were accepted during this period.

\subsubsection{Tree Depth, Node Number, and Split Distribution}
\label{sec:treeDepth}
\begin{figure}[!b]
\begin{center}
   \begin{subfigure}{0.49\linewidth} \centering
     \includegraphics[scale=0.3]{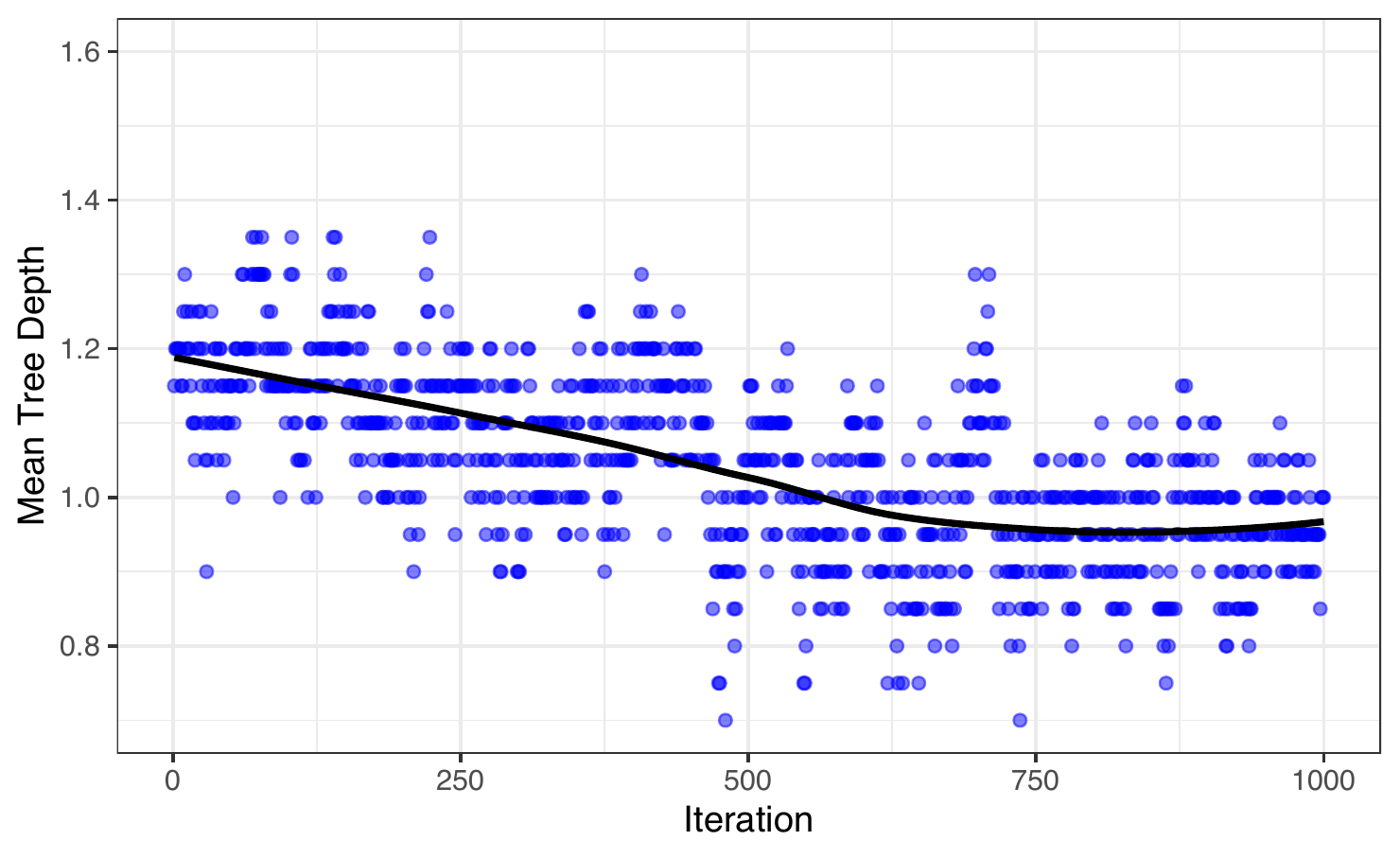} 
     \captionsetup{justification=centering}
     \caption{Average tree depth from \texttt{bartMachine}.}
     \label{}
   \end{subfigure}
   \begin{subfigure}{0.49\linewidth} \centering
     \includegraphics[scale=0.3]{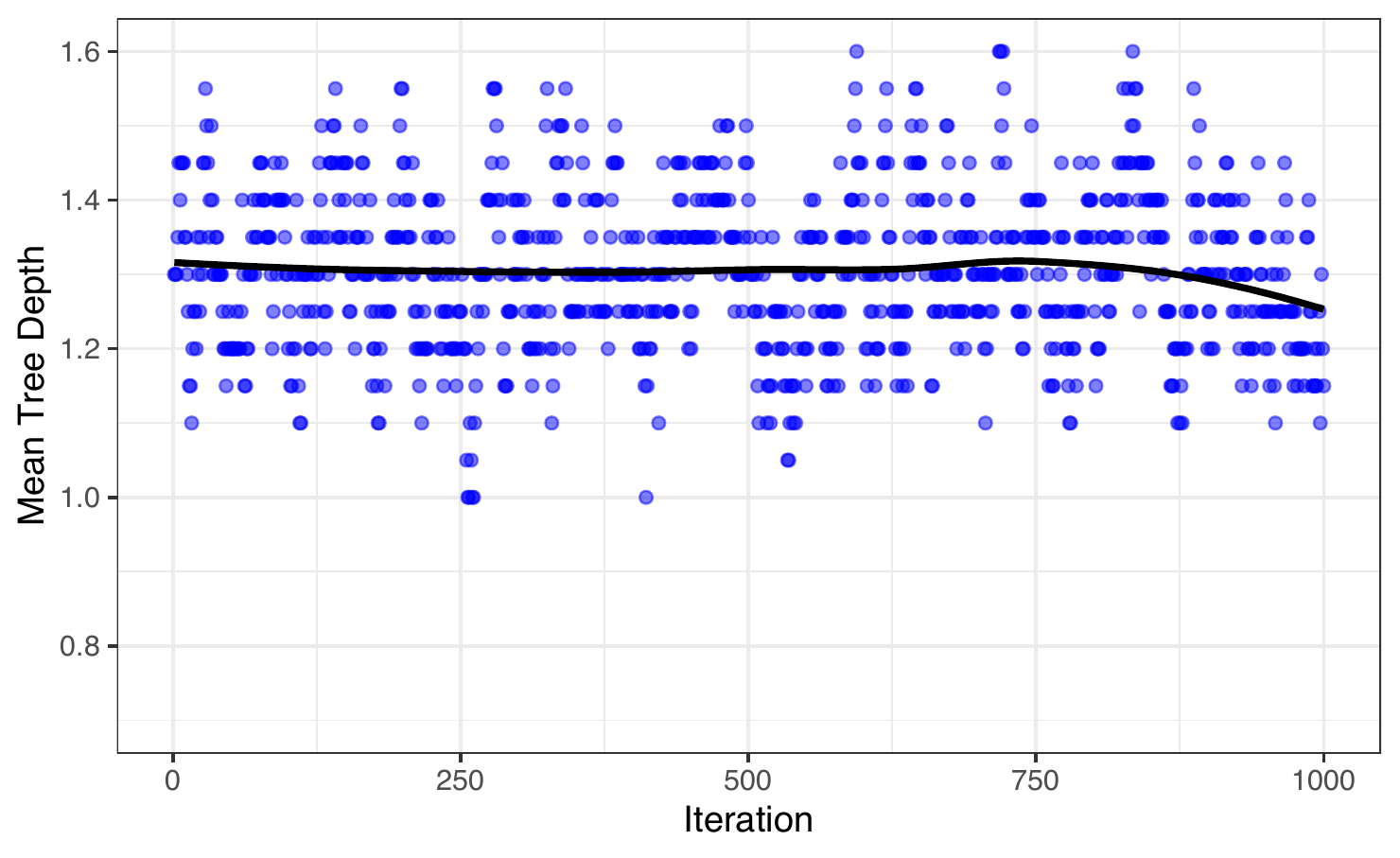} 
     \captionsetup{justification=centering}
     \caption{Average tree depth from \texttt{dbarts}.}\label{}
   \end{subfigure}
    \begin{subfigure}{0.49\linewidth} \centering
     \includegraphics[scale=0.3]{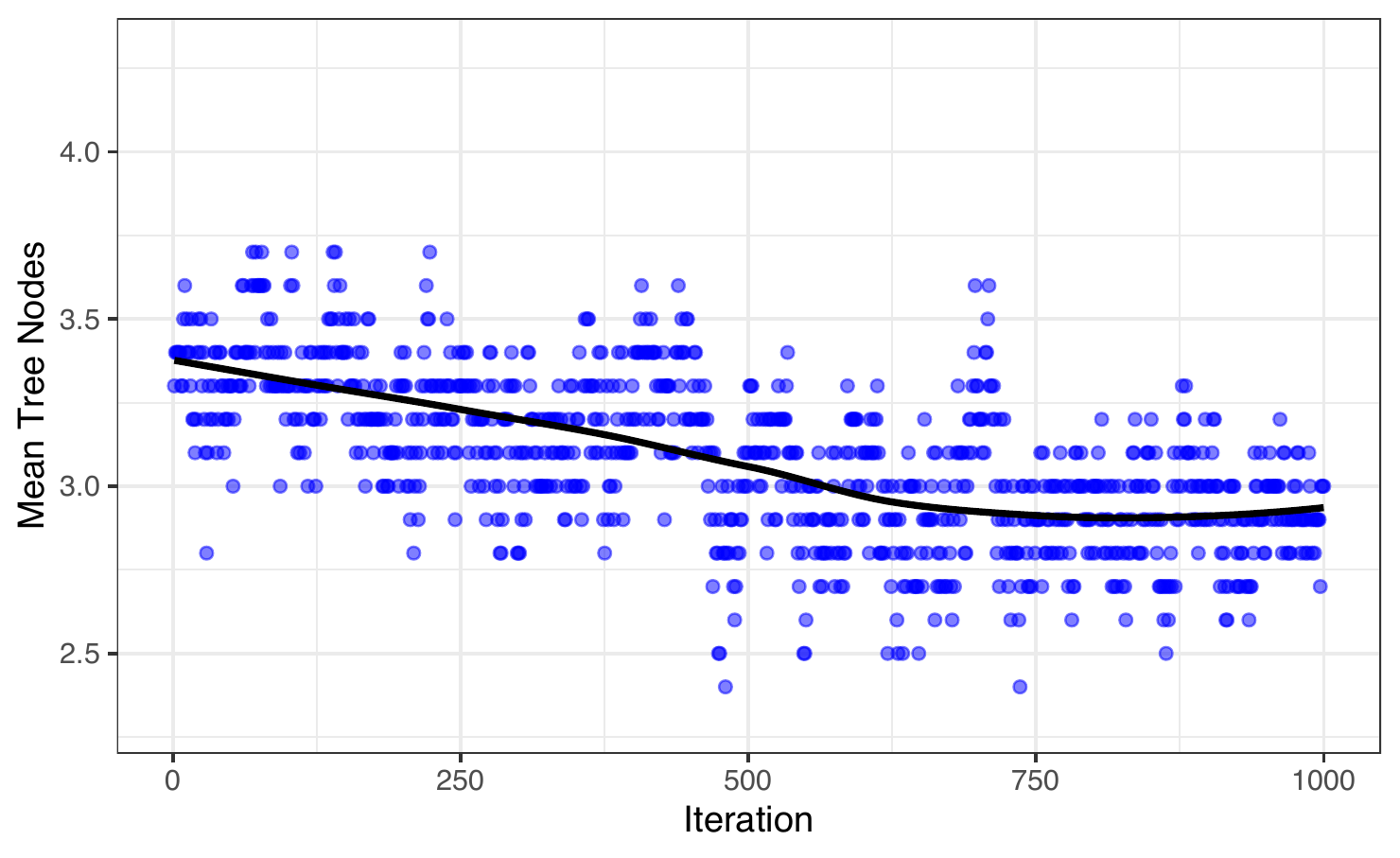} 
     \captionsetup{justification=centering}
     \caption{Average number of nodes from \texttt{bartMachine}.}\label{}
   \end{subfigure}
    \begin{subfigure}{0.49\linewidth} \centering
     \includegraphics[scale=0.3]{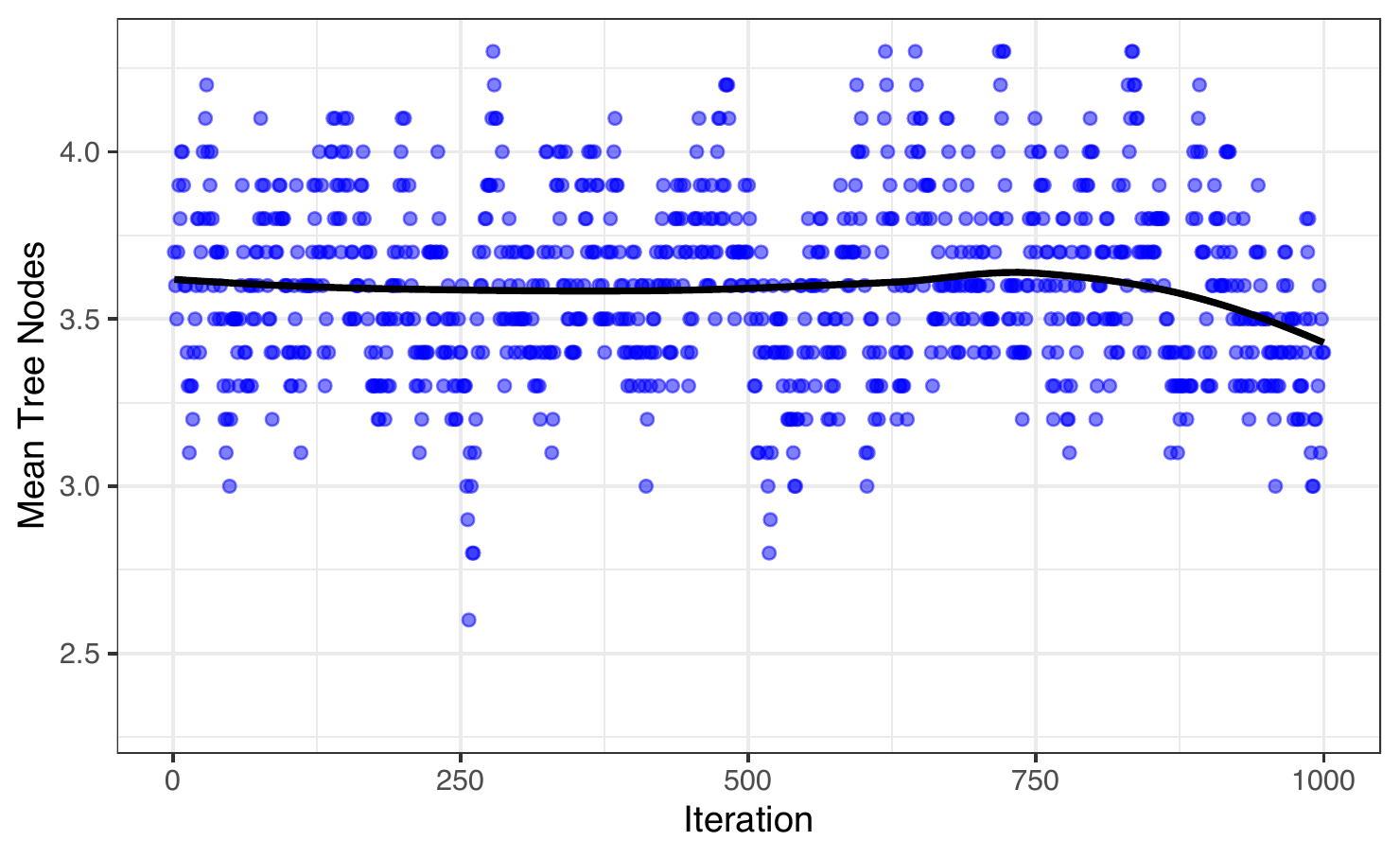}
     \captionsetup{justification=centering}
     \caption{Average number of nodes from \texttt{dbarts}.}\label{}
   \end{subfigure}
\caption{In the top row we show the post burn-in average tree depth per iteration for a \texttt{bartMachine} and \texttt{dbarts} fit in (a) and (b), respectively. In the bottom row we show the post burn-in average number of nodes per iteration for a \texttt{bartMachine} and \texttt{dbarts} fit in (c) and (d), respectively. A black LOESS regression curve is shown to indicate the changes in both the average tree depth and number of nodes across iterations.} 
\label{fig:depthAndNodes}
\end{center}
\end{figure}

As with the acceptance rate, the average tree depth and average number of all nodes per iteration can give an insight into the fit's stability. Figure \ref{fig:depthAndNodes} 
displays these two metrics for both BART fits.  A locally estimated scatterplot smoothing (LOESS) regression line is shown to indicate the changes in both the average tree depth and the average number of nodes across iterations.
From Figure \ref{fig:depthAndNodes} (a) and (c), we can see that both the post burn-in average tree depth and the average number of nodes per iteration is much more stable in the \texttt{dbarts} fit. However, although we use the default number of iterations suggested by the \texttt{bartMachine} package, increasing this may improve stability.

\begin{figure}[!t]
\begin{center}
   \begin{subfigure}{0.49\linewidth} \centering
     \includegraphics[scale=0.45]{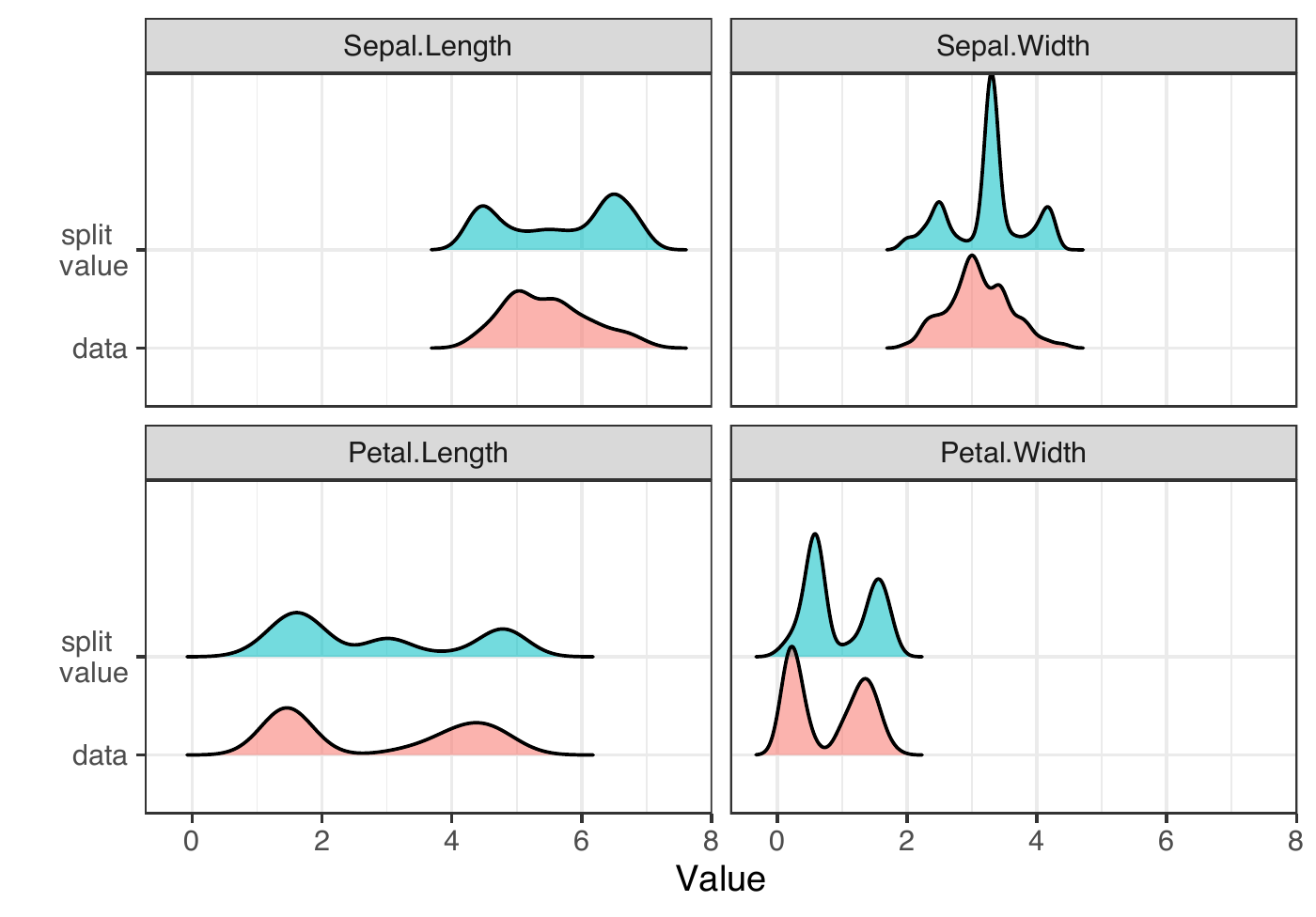} 
     \captionsetup{justification=centering}
     \caption{Split value distribution obtained from a \texttt{bartMachine} fit.}
   \end{subfigure}
   \begin{subfigure}{0.49\linewidth} \centering
     \includegraphics[scale=0.45]{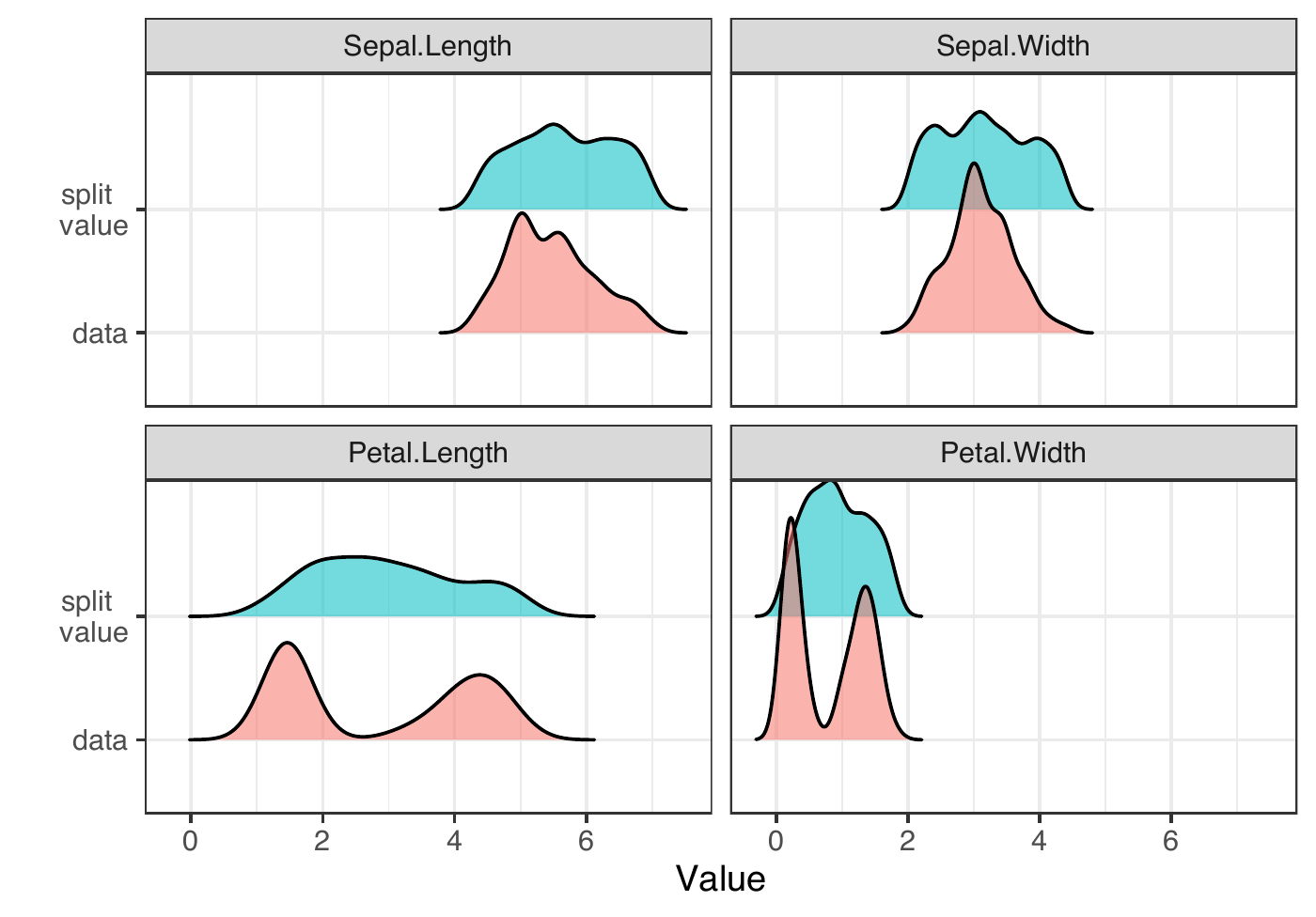} 
     \captionsetup{justification=centering}
     \caption{Split value distribution obtained from a \texttt{dbarts} fit.}
   \end{subfigure}
\caption{Split values densities (in green) over all iterations for each variable overlayed on the densities of the predictors (in red) for a \texttt{bartMachine} fit in (a) and a \texttt{dbarts} fit in (b).} 
\label{fig:toySplitDens}
\end{center}
\end{figure}

Figure \ref{fig:toySplitDens} 
shows the densities of split values over all post burn-in iterations for each variable for both models (in green), combined with the densities of the predictor variables (labeled ``data'', in red). This plot appears to be new; we have not found anything similar in any of the existing packages.  We can see that the split value density for Sepal.Width in the \texttt{bartMachine} fit, in (a), has large peak at around 3.2 and the  \texttt{bartMachine} fit's split values have more modes.

In addition to the previous plots, we provide a panel of basic summary diagnostics of the model fit which can be used for both classification and regression models. For the former, we display metrics such as precision-recall and ROC (with uncertainties included), a confusion matrix, fitted value plots, and a histogram of predicted probabilities. For the latter, we show a trace plot of the model variance, a Q-Q plot, and an array of model performance plots and residual plots over all iterations. In the interest of space, we exclude the summary diagnostics for the classification model and display the summary diagnostic plots for the regression model only, as seen in Section \ref{sec:caseStudy}.

\section{Comparative analysis of variable Importance and Interactions in a BART model}
 \label{sec:suppMat}

In this section we provide an examination of the variable inclusion proportion methods for evaluating importance and interactions in a BART model (as outlined in Section \ref{sec:viviBARTintro}) by comparing the raw inclusion proportions with and without uncertainty included against alternative methods used to assess the importance and interactions of variables. These alternative methods do not allow for the inclusion of uncertainty in the metrics they create. 

As previously discussed, BART models obtain a measure of importance by observing the proportion of times a variable is used as a split variable across all trees, averaged over all iterations. The more times a variable is used as a split variable, the more important that variable is deemed to be. Similarly, a measure of interaction can be obtained in a BART model by observing the proportion of successive splits over all trees, averaged over all iterations. However, as noted by \cite{chipman2010bart}, this method of assessing importance (and interactions) comes with certain pitfalls. Namely, if the number of trees is large, then non-important predictor variables can be preferred as the likelihood is relatively flat and so the tree prior dominates. This can lead to to spurious importance and interactions scores for variables that, in reality, have little influence on the response. This effect can be mitigated somewhat by the inclusion of uncertainty to evaluate the reliability of the measured importance or interaction scores. Additionally,  \cite{chipman2010bart} state that decreasing the number of trees when building the model diminishes this effect as less important variables get swapped out of the trees for more informative variables.
 
To compare the usefulness of a BART model's importance and interactions, we compare the BART methodology, with and without uncertainty included, against a model agnostic approach to assess the importance and interactions. To measure the agnostic variable importance we use a permutation method. Permutation importance was first introduced by \cite{breiman2001randomforest} and works by calculating the change in the model's predictive performance after a variable has been randomly permuted. That is, a model score is initially recorded, then a single variable is randomly permuted (this is repeated for each variable) and the model score is recalculated on the new dataset. The difference between the baseline model's performance and the permuted model's performance is taken as the variable importance score. To measure the agnostic interactions we use Friedman's $H$-statistic (or $H$-index) \citep{Friedmans_H}. For this method the partial dependence for a pair of variables is compared to their marginal effects. 

Friedman's $H$-statistic is defined as:
\begin{equation}
H^2_{jk} = \frac{\sum_{i=1}^{n}[f_{jk}(x_{ij}, x
_{ik}) - f_{j}(x_{ij}) - f_{k}(x_{ik})]^2 }{\sum_{i=1}^{n}f^2_{jk}(x_{ij}, x_{ik})}
\label{eqn:HStatInPaper}
\end{equation}
\noindent where $f_{jk}(x_j,x_k)$ represents the two-way partial dependence function of both variables, $f_{j}(x_j)$ and $f_{k}(x_k)$ represent the partial dependence functions of the single variables, and all partial dependence functions are mean-centered. The obtained measure is scaled in the range (0,1). \cite{inglis2022visualizing} note, however, that variations in the numerator can lead to spuriously high $H$-values when the denominator in (\ref{eqn:HStatInPaper}) is small because the partial dependence function for the variables $j$ and $k$ is flat in this case. To combat this, the square-root of the average un-normalized (numerator only) version of Friedman's $H^2$ for calculating pairwise interactions is suggested:
\begin{equation}
H_{jk} = \sqrt {  \frac{1}{n} \sum_{i=1}^{n}[f_{jk}(x_{ij}, x
_{ik}) - f_{j}(x_{ij}) - f_{k}(x_{ik})]^2   }
\label{eqn:HStatInPapernn}
\end{equation}

To explore and compare the variable importance and variable interactions, we generate data using the Friedman benchmark equation \citep{FriedmanEqn}:
\begin{equation}
\begin{aligned}
y = 10 \sin(\pi x_1 x_2) + 20 (x_3 - 0.5)^2 + 10 x_4 + 5 x_5 + \epsilon
\\{\rm where} \;\; x_j \sim  U(0, 1), j=1,2,\ldots, 10;\;\;   \epsilon \sim N(0, 1). \notag
\label{eqn:friedData}
\end{aligned}
\end{equation}
We simulate 250 observations and fit a BART model using the \texttt{dbarts} R package, using the default number of iterations (1000) and burn-in (100). We then set the number of trees to be 20, 100, and 200 to evaluate how well the BART model can capture the importance and interactions. There are five important variables and an interaction between $x_1$ and $x_2$ in Equation \ref{eqn:friedData}, and five additional predictors $x_6, x_7, \ldots x_{10}$ unrelated to the response.

\begin{figure}[!t]
\begin{center}
   \begin{subfigure}{0.32\linewidth} \centering
     \includegraphics[scale=0.23]{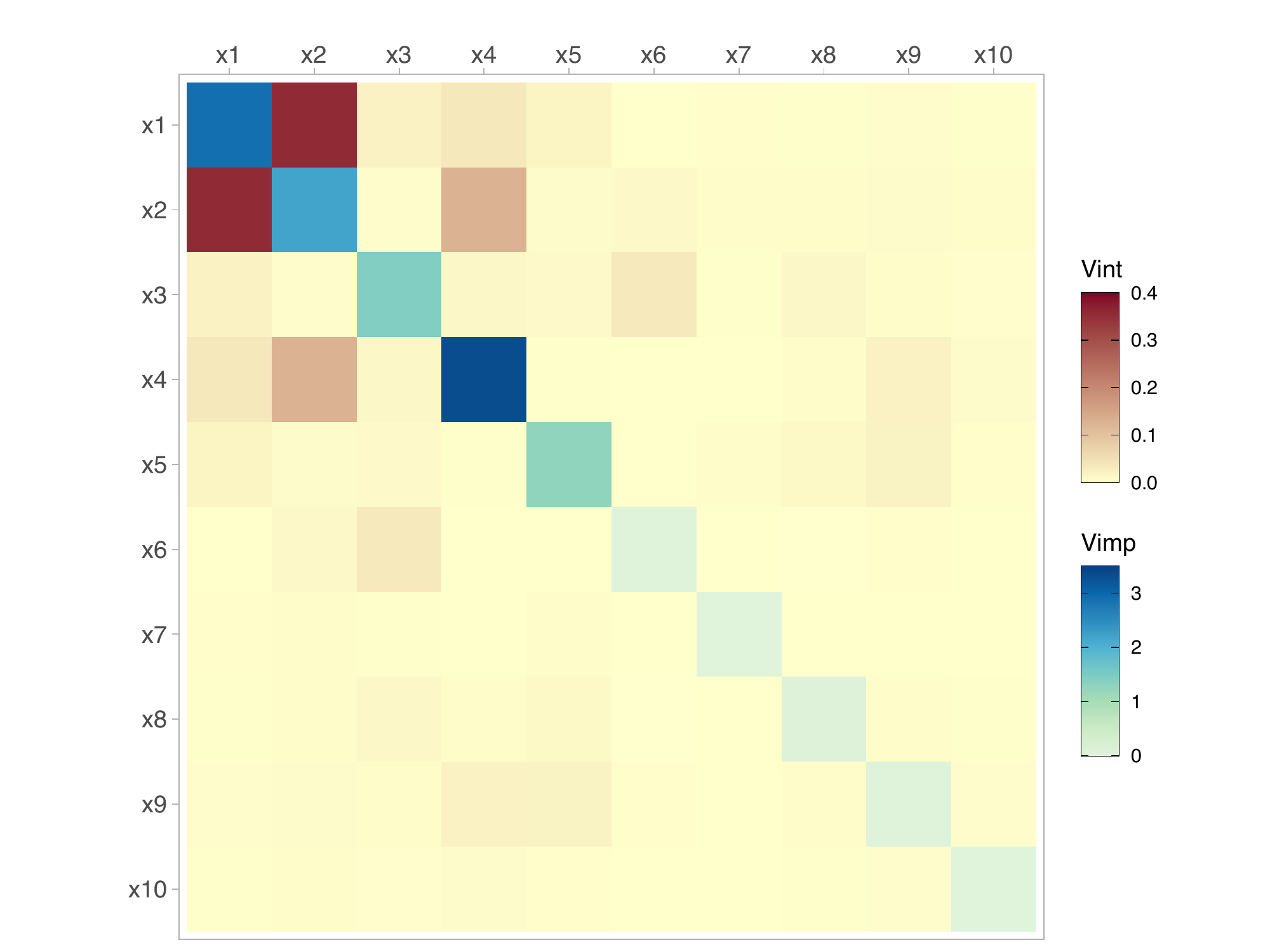} 
     \captionsetup{justification=centering}
     \caption{Agnostic Method \\with 20 trees.}
   \end{subfigure}
   \begin{subfigure}{0.32\linewidth} \centering
     \includegraphics[scale=0.23]{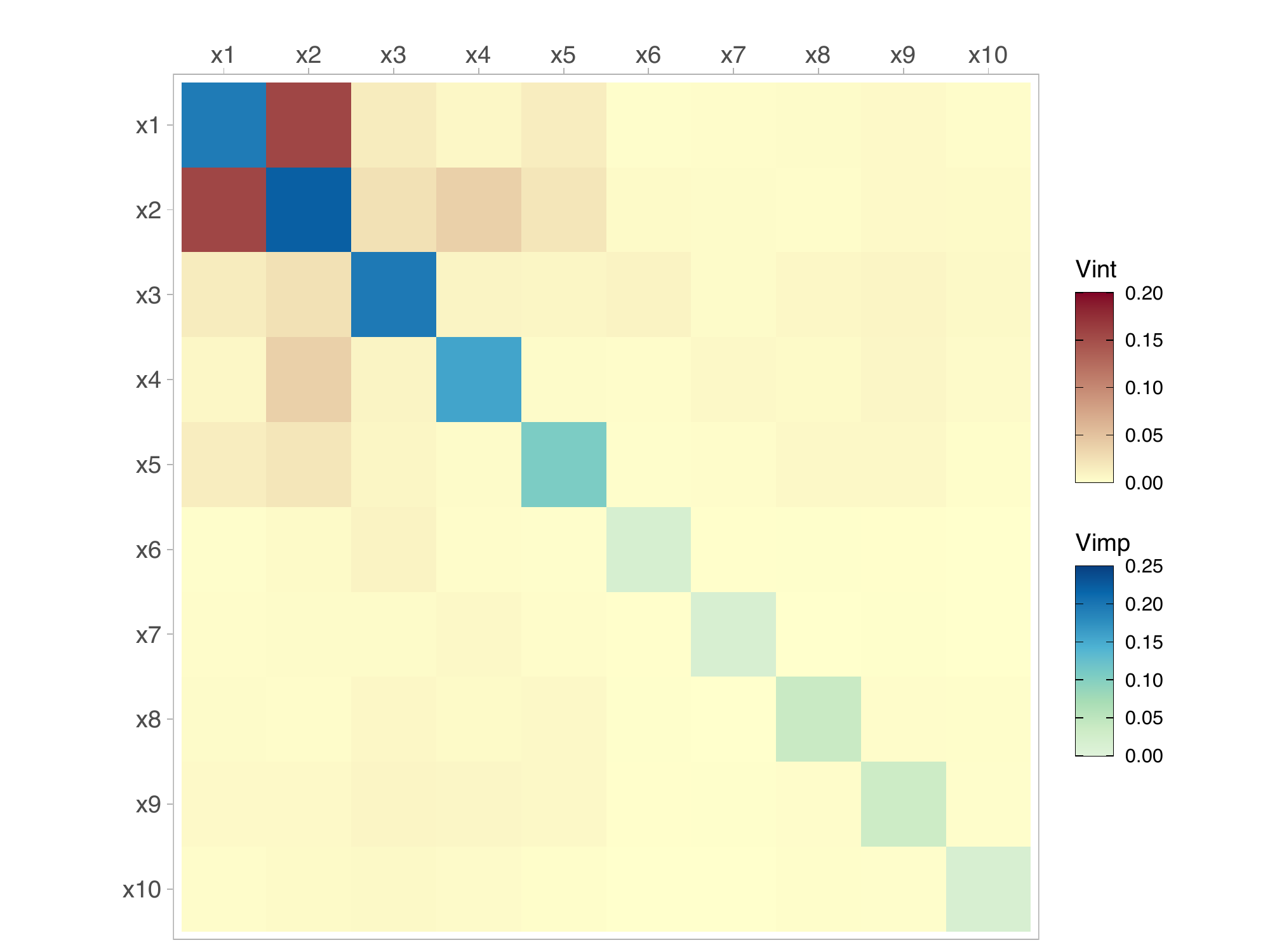} 
     \captionsetup{justification=centering}
     \caption{Inclusion proportion method with 20 trees.}
   \end{subfigure}
   \begin{subfigure}{0.32\linewidth} \centering
     \includegraphics[scale=0.23]{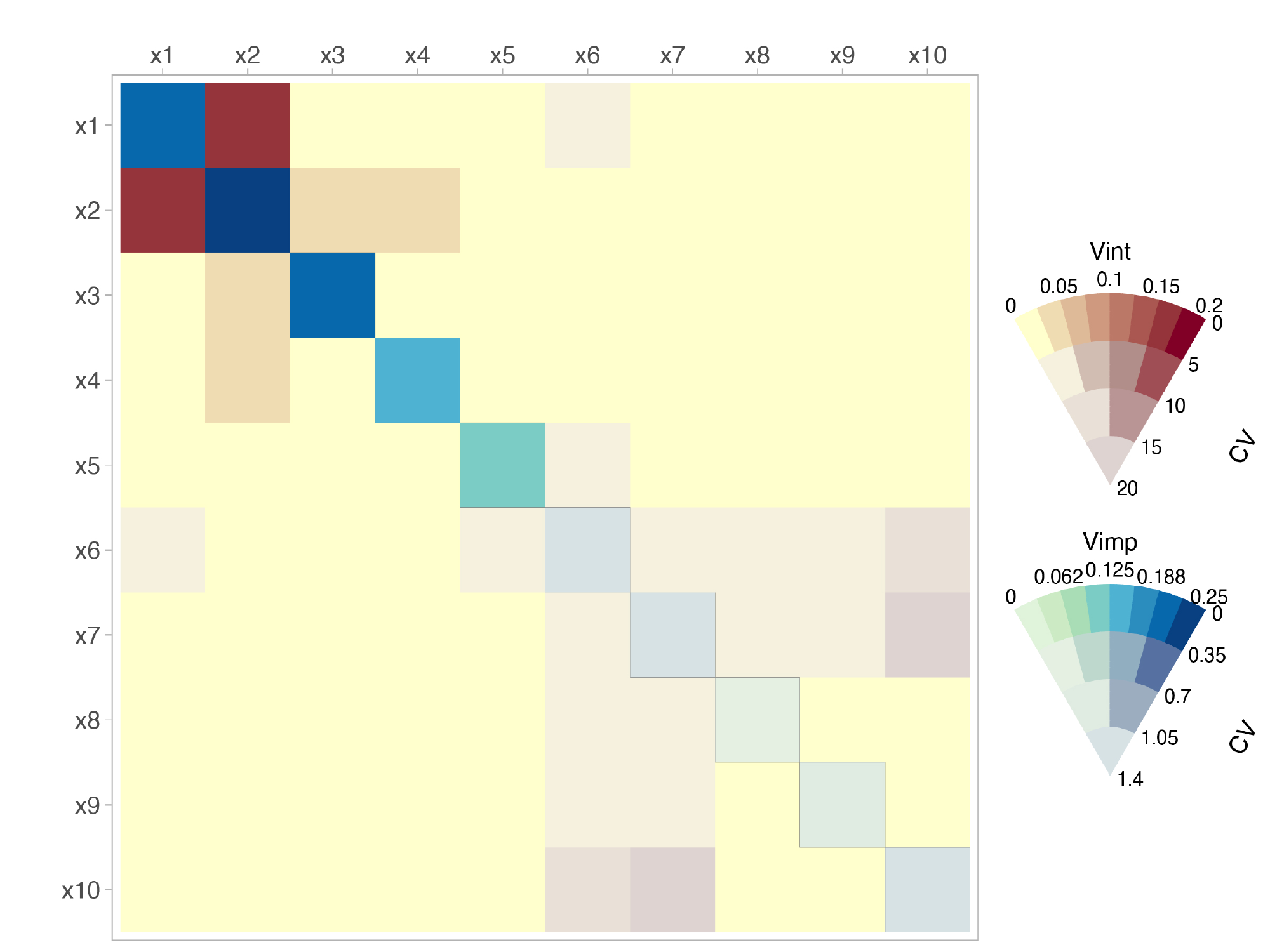}
     \captionsetup{justification=centering}
     \caption{Inclusion proportion with uncertainty with 20 trees.}
   \end{subfigure}
    \begin{subfigure}{0.32\linewidth} \centering
     \includegraphics[scale=0.23]{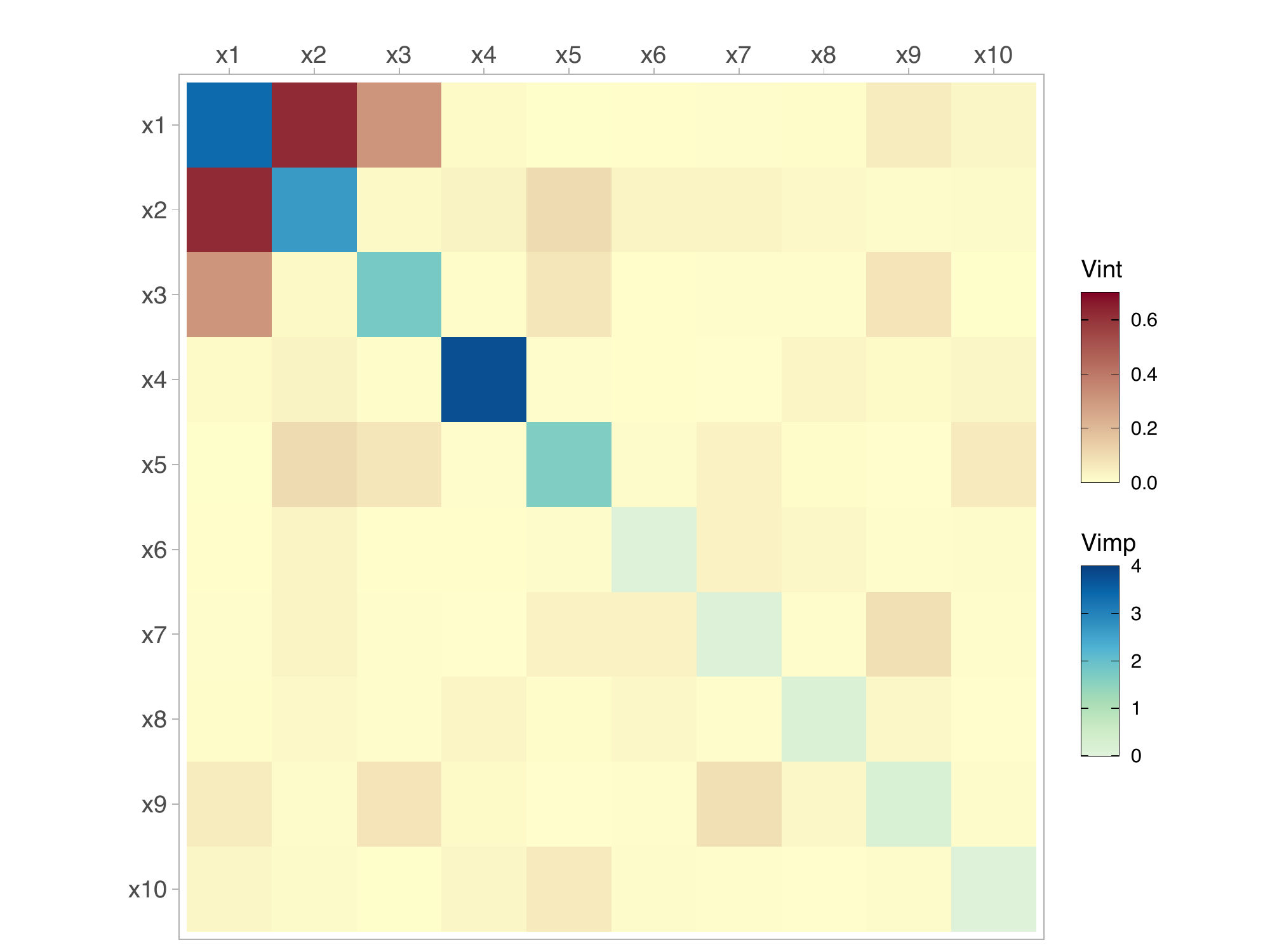} 
     \captionsetup{justification=centering}
     \caption{Agnostic Method \\with 100 trees.}
   \end{subfigure}
   \begin{subfigure}{0.32\linewidth} \centering
     \includegraphics[scale=0.23]{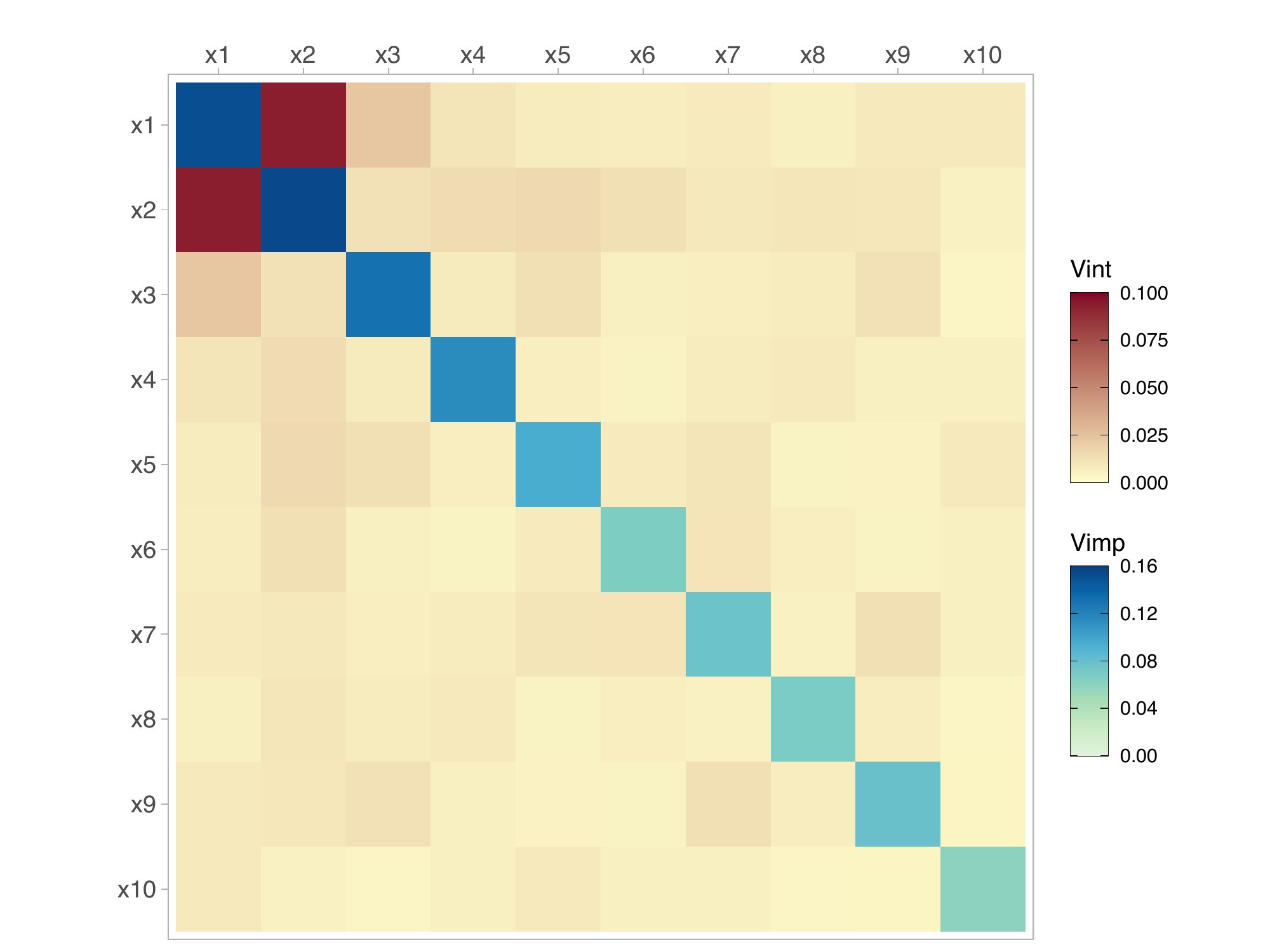} 
     \captionsetup{justification=centering}
     \caption{Inclusion proportion method with 100 trees.}
   \end{subfigure}
   \begin{subfigure}{0.32\linewidth} \centering
     \includegraphics[scale=0.23]{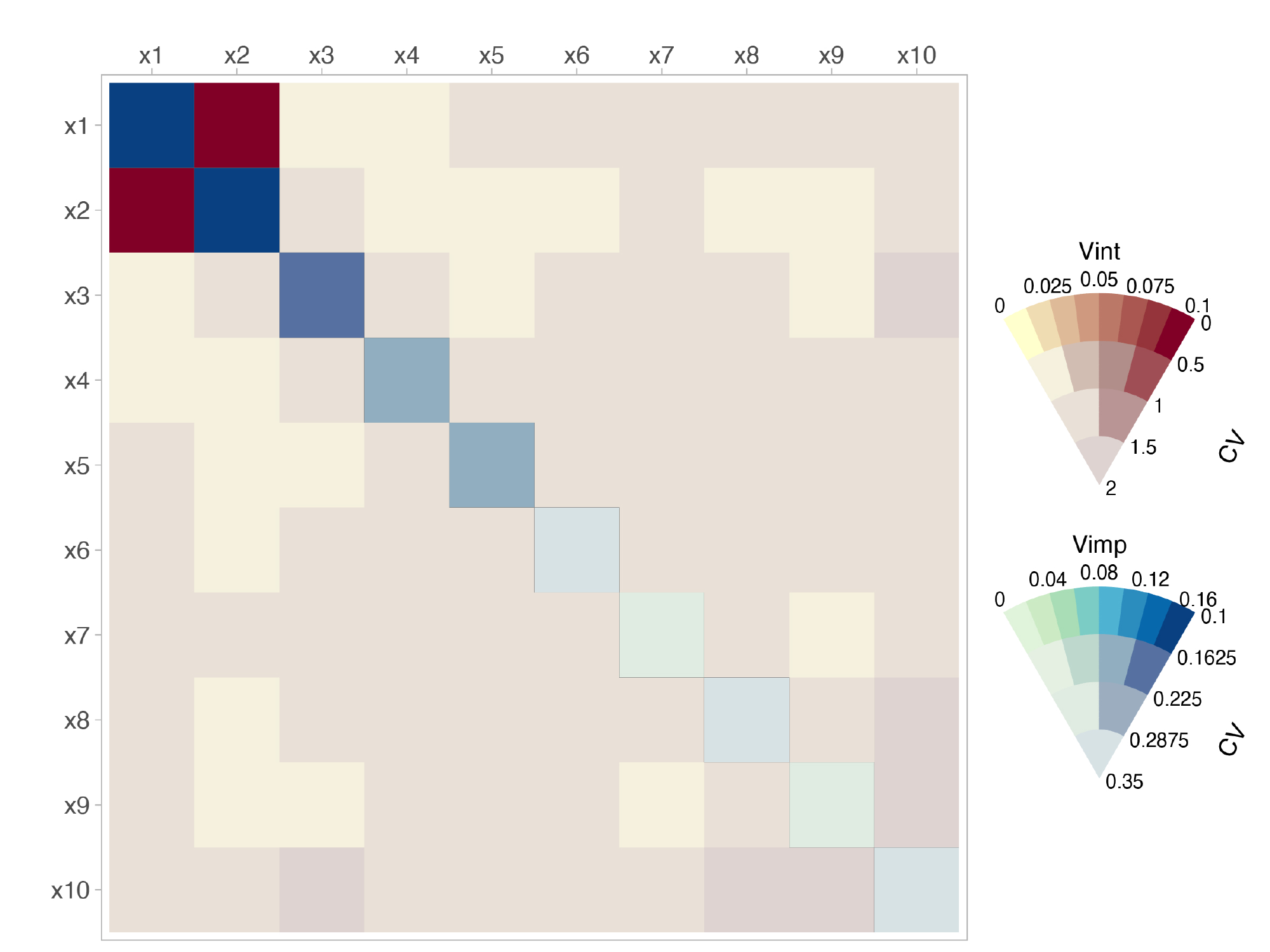}
     \captionsetup{justification=centering}
     \caption{Inclusion proportion with uncertainty with 100 trees.}
   \end{subfigure}
    \begin{subfigure}{0.32\linewidth} \centering
     \includegraphics[scale=0.23]{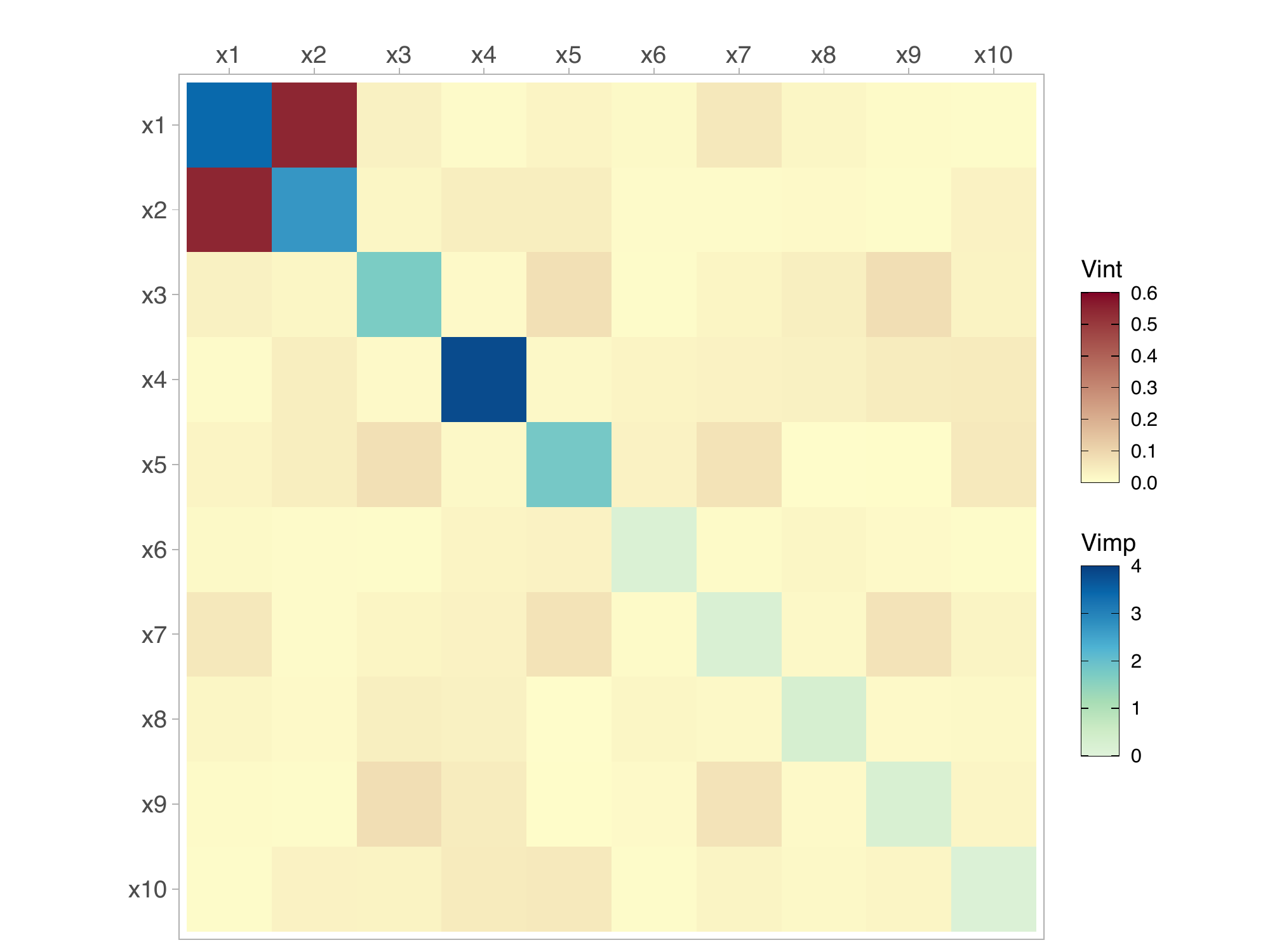} 
     \captionsetup{justification=centering}
     \caption{Agnostic Method \\with 200 trees.}
   \end{subfigure}
   \begin{subfigure}{0.32\linewidth} \centering
     \includegraphics[scale=0.23]{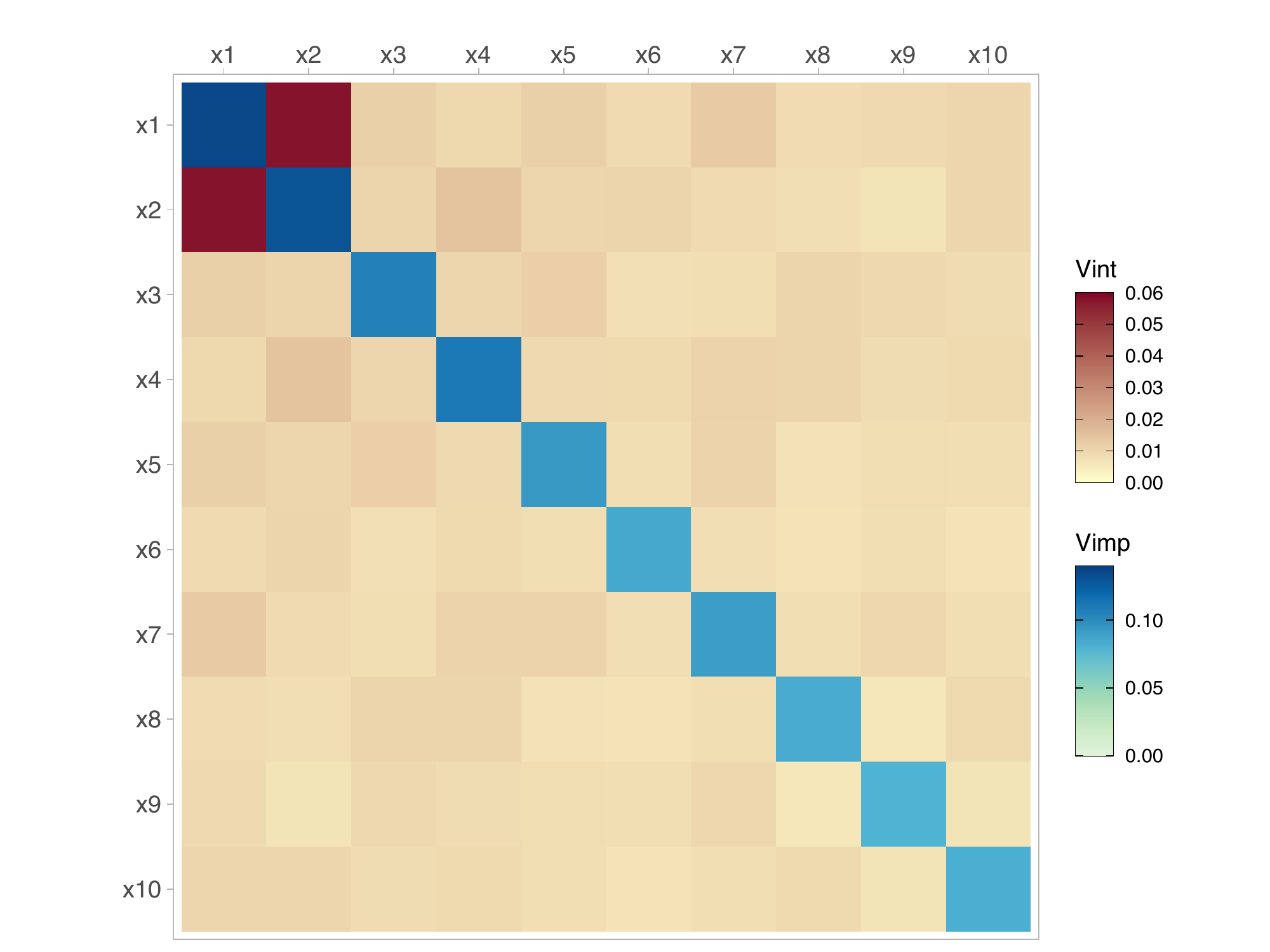} 
     \captionsetup{justification=centering}
     \caption{Inclusion proportion method with 200 trees.}
   \end{subfigure}
   \begin{subfigure}{0.32\linewidth} \centering
     \includegraphics[scale=0.23]{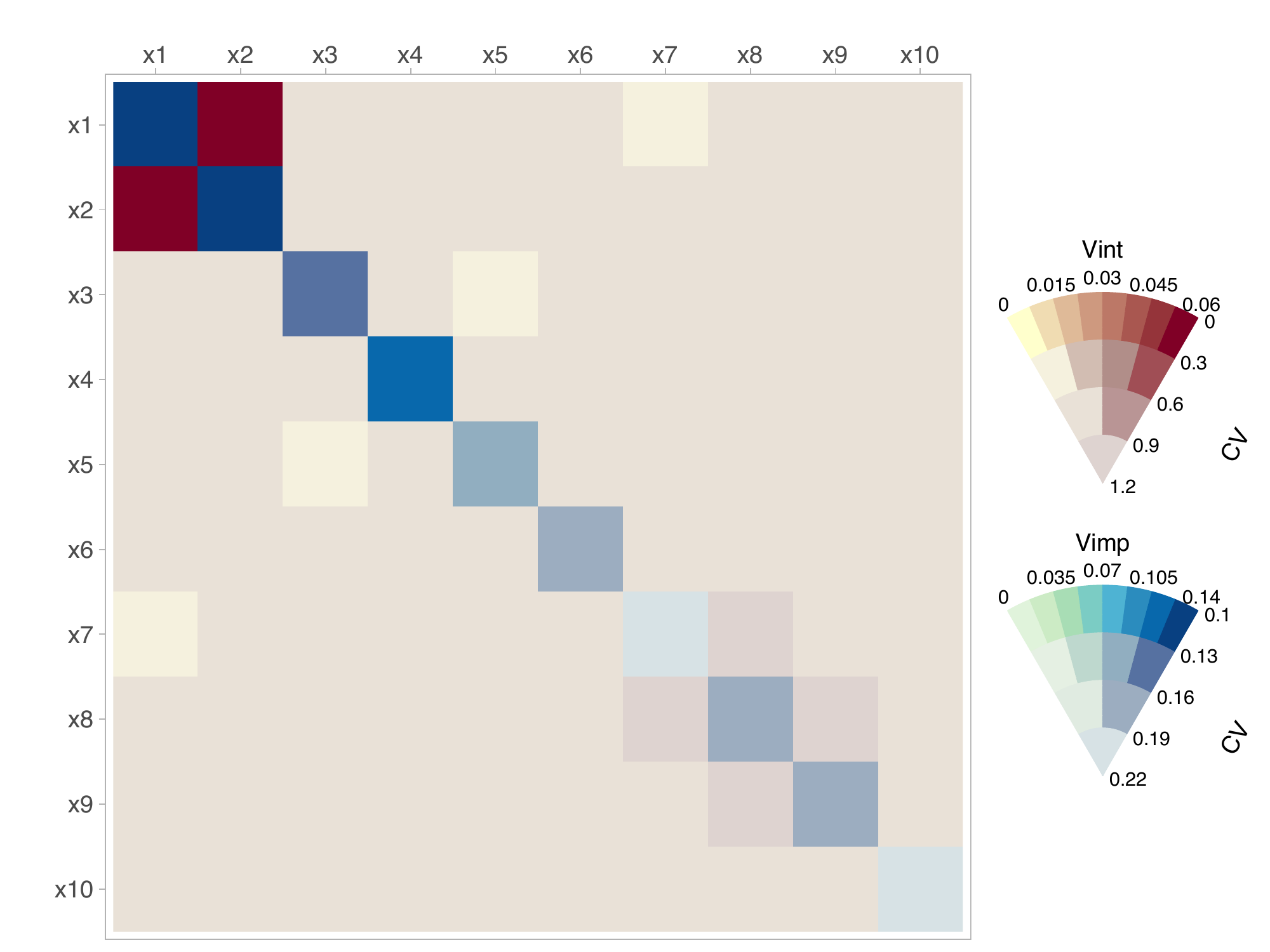}
     \captionsetup{justification=centering}
     \caption{Inclusion proportion with uncertainty with 200 trees.}
   \end{subfigure}
   \caption{Comparison of different methods to determine importance and interactions in a BART model with 20, 100, and 200 trees in the first, second, and third rows respectively.}
\label{fig:comparison}
\end{center}
\end{figure}

In the first column of Figure \ref{fig:comparison} 
(panels (a), (d), and (g)) we use the alternative agnostic permutation approach for measuring importance and Friedman's $H$-statistic to obtain the interaction measures. In the second column (panels (b), (e), and (h))  we calculate the standard BART model variable inclusion proportion for the importance and interactions. Finally, in the third column (panels (c), (f), and (i)), we display the same information as in the second column but with uncertainty included, in this case via the coefficient of variation. For each row of Figure \ref{fig:comparison} we set the number of trees to 20, 100, and 200.

Using the alternative agnostic method (first column) the five important variables are identified with $x_4$ being ranked as the most important and the interaction between $x_1$ and $x_2$ is prominent. This remains consistent, regardless of the number of trees used when building the model. When using the inclusion proportions (second column) the interaction between  $x_1$ and $x_2$ is strong and individually $x_1$ and $x_2$ are the most important. In (b) the five important variables are identified. However, as the number of trees increases (see (e) and (h)) variables $x_6, \ldots, x_{10}$ are incorrectly  designated as important. Spurious values are measured for both  importance and interactions when increasing the number of trees. Examining the VSUPs (third column)  the interaction between $x_1$ and $x_2$ is prominent and the five important variables are again evident. Increasing the number of trees has the effect of increasing the relative uncertainty for the spurious values and therefore, highlights the variables of interest. For example, if we compare panels (e) and (f) each based on 100 trees, we see that most of the spurious importance and interaction values in (e) have a moderate degree of relative uncertainty in (f).

It is worth noting that for 20, 100, or 200 trees, although the agnostic method had relatively consistent results, this method may not be computationally practical as it is a slow calculation which gets compounded by the increase in trees. Additionally, the agnostic approach would have be repeated multiple times to allow a measure of uncertainty to be obtained. Conversely, calculating the inclusion proportion is quick. For example, calculating the inclusion proportion for importance and interactions for when the number of trees is 20 (as in panels (b) and (c)) took approximately 1.5 seconds on a MacBook Pro 2.3 GHz Dual-Core Intel Core i5 with 8GB of RAM. Whereas, using the agnostic approach to measure the importance and uncertainty (as in panel (a)) took approximately 43 seconds on the same machine. When viewed with the uncertainty included, the inclusion proportion method performs well when compared to the agnostic method, particularly when the number of trees is low.

\section{Case Study: Seoul Bike Sharing Data}
\label{sec:caseStudy}

In this section we apply our methods on a larger real-world data set. Here we examine and create visualizations concerning bike sharing data from Seoul, South Korea \citep{bike}. The data contains 14 features and includes weather data (for example, humidity, rainfall, snowfall, and several others), the time of the bike rental (in seasons, months, and days), and some local information (such as if the day of rental was a holiday), with the total number of bikes rented per day as the response. The original data contained 8760 hourly observations which we summarize to obtain the daily counts. For a full description of the data see the Supplementary Materials. The data has been previously studied in \cite{ve2020rule} and \cite{sathishkumar2020using} who found that the temperature of the day was an important factor for predicting the total number of rentals. \cite{ve2020season} also found that  the individual month and season  play a significant role in predicting bike rentals. 

For our study we fit a BART model, using the \texttt{BART} package, with 1000 iterations, a burn-in of 100, and 100 trees, with the goal of investigating which of the predictor variables has a significant impact on the response. We apply a cube root transformation to the response as initially the residuals displayed some evidence of  non-normality. As mentioned in Section \ref{sec:viviBARTintro} on factor dummy variables, we perform an aggregation of the dummy variables’ inclusion proportions for both the importance and the interactions so these metrics can be assessed on the entire factor. The variables treated as factors in the data are Month (the month of the year a bike is rented), Season (season of the year a bike is rented), Wkend (if the day of bike rental is a weekend or not), and Holiday (if the day of bike rental is a public holiday or not).

\begin{figure}[!b]
\centering
\includegraphics[scale = 0.5]{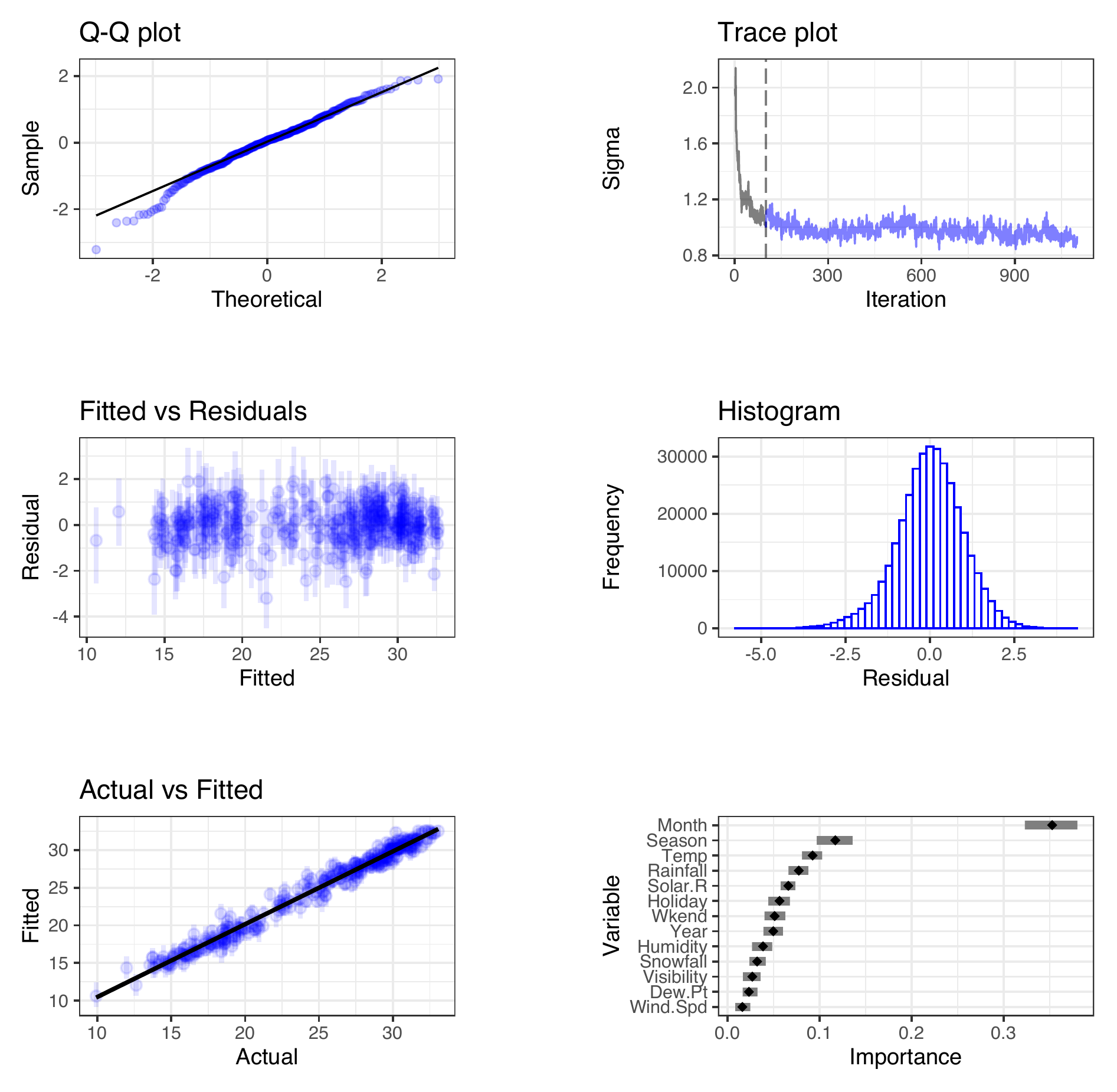}
\caption{General diagnostic plots for a BART regression fit on bike sharing demand data. Top left: A QQ-plot of the residuals after fitting the model. Top right: $\sigma$ by MCMC iteration. Middle left: Residuals versus fitted values with 95\% credible intervals. Middle right: A histogram of the residuals. Bottom Left: Actual values versus fitted values with 95\% credible intervals. Bottom right: Variable importance plot with 25 to 75\% quantile interval shown. We can see in the bottom left panel that the model fits the training data reasonably well, with a good convergence seen in the top right panel. }
\label{fig:bsdiagRegression}
\end{figure}

To begin Figure \ref{fig:bsdiagRegression} 
shows the model's diagnostics to assess the stability of the model fit. The top two rows indicate a reasonable performance of the residuals with a moderately stable convergence of the residual standard deviation. The black vertical line in the trace plot indicates the separation between the pre and post burn-in period. The bottom row shows that the model fits the training data well and that the Month is clearly the most important variable for predicting the count of bikes rented. However, in the bottom right panel, we can see that Month has a large 25-75\% quantile interval when compared to the other variables. The second most important variable is Season followed by Temp (average daily temperature in $^{\circ}C$).

\begin{figure}[!b]
\begin{center}
   \begin{subfigure}[t]{0.49\linewidth} \centering
     \includegraphics[scale=0.35]{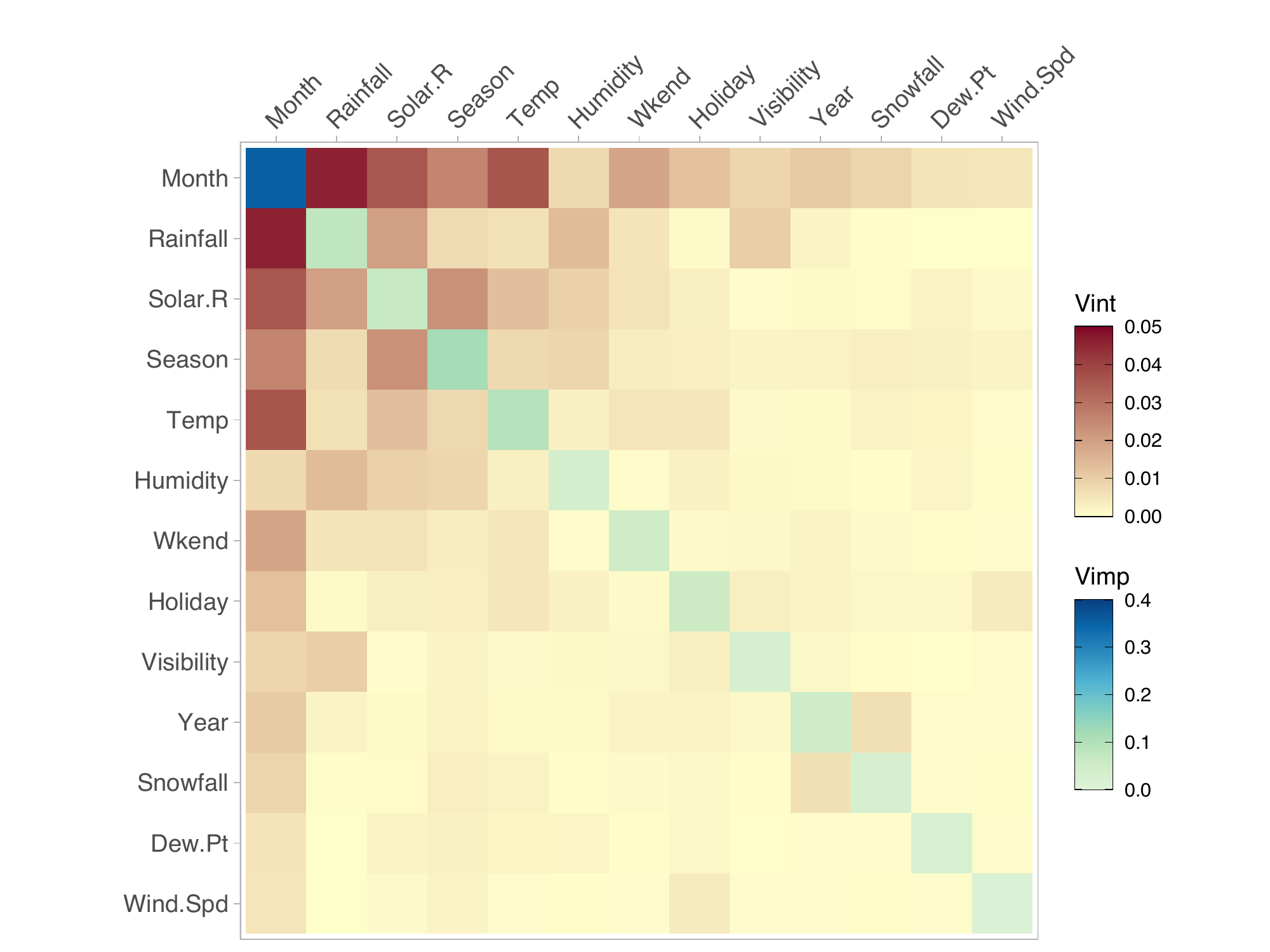} 
     \captionsetup{justification=centering}
     \caption{}
   \end{subfigure}
   \begin{subfigure}[t]{0.49\linewidth} \centering
     \includegraphics[scale=0.35]{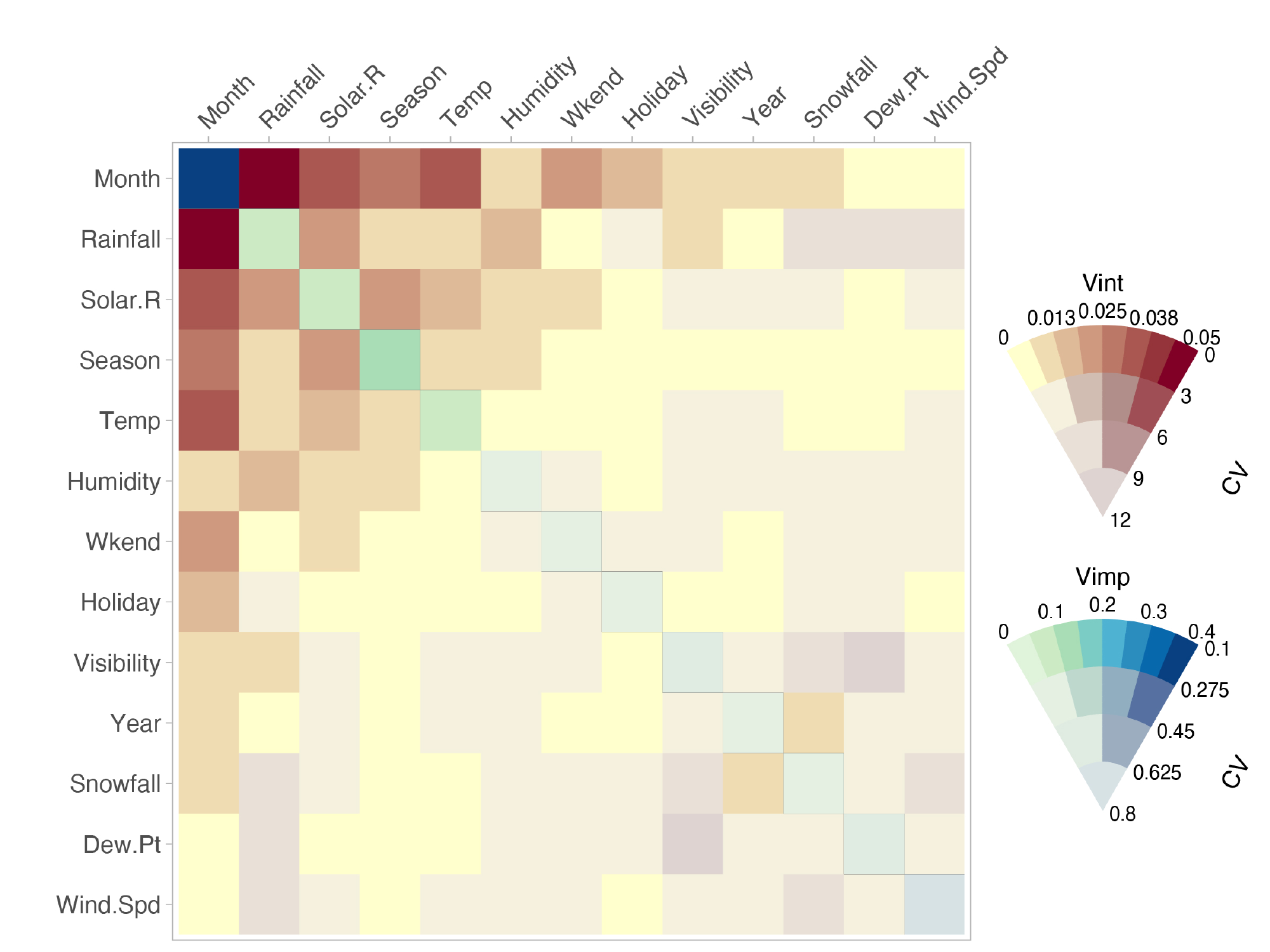} 
     \captionsetup{justification=centering}
      \caption{}
   \end{subfigure}
\caption{In (a) Variable importance and interaction plot without uncertainty.  In (b) the same values  are shown but with the uncertainty
included by use of a VSUP. In (b) we can see that the interaction values between Month and several other variables have a low coefficient of variation associated with them.}
\label{fig:vivibs}
\end{center}
\end{figure}

We explore the impact of the variables on the response by examining the importance and interactions jointly in the variable importance and variable interaction plots of Figure \ref{fig:vivibs}. 
For illustration purposes only we show the plot without uncertainty in the left panel and with uncertainty on the right (as before we use the coefficient of variation). In Figure \ref{fig:vivibs} (a) we observe a strong interaction between the variable Month and several others, notably; Rainfall (in mm), Solar.R (Solar radiance in $mJ/m^2$), Season, and Temp. The strongest interaction can be seen between Month and Rainfall. In Figure \ref{fig:vivibs}(b) many of the low importance  and interaction scores have high relative uncertainty, so the viewer's attention is drawn to the interesting variables. The most important variable Month remains important relative to its uncertainty.  Equally, the strong interactions observed in (a) between Month and several others have a low associated variation in (b). In (a) all variables except Month have similar importance scores, but relative to uncertainty, the importance of the last seven variables (Humidity to Wind.Spd) is reduced, represented by greeny-grey colors along the diagonal (b). The interactions between these variables are mostly low and/or with high relative uncertainty, the interaction between Snowfall and Year being an exception.

\begin{figure}[!t]
\centering
\includegraphics[scale = 0.5]{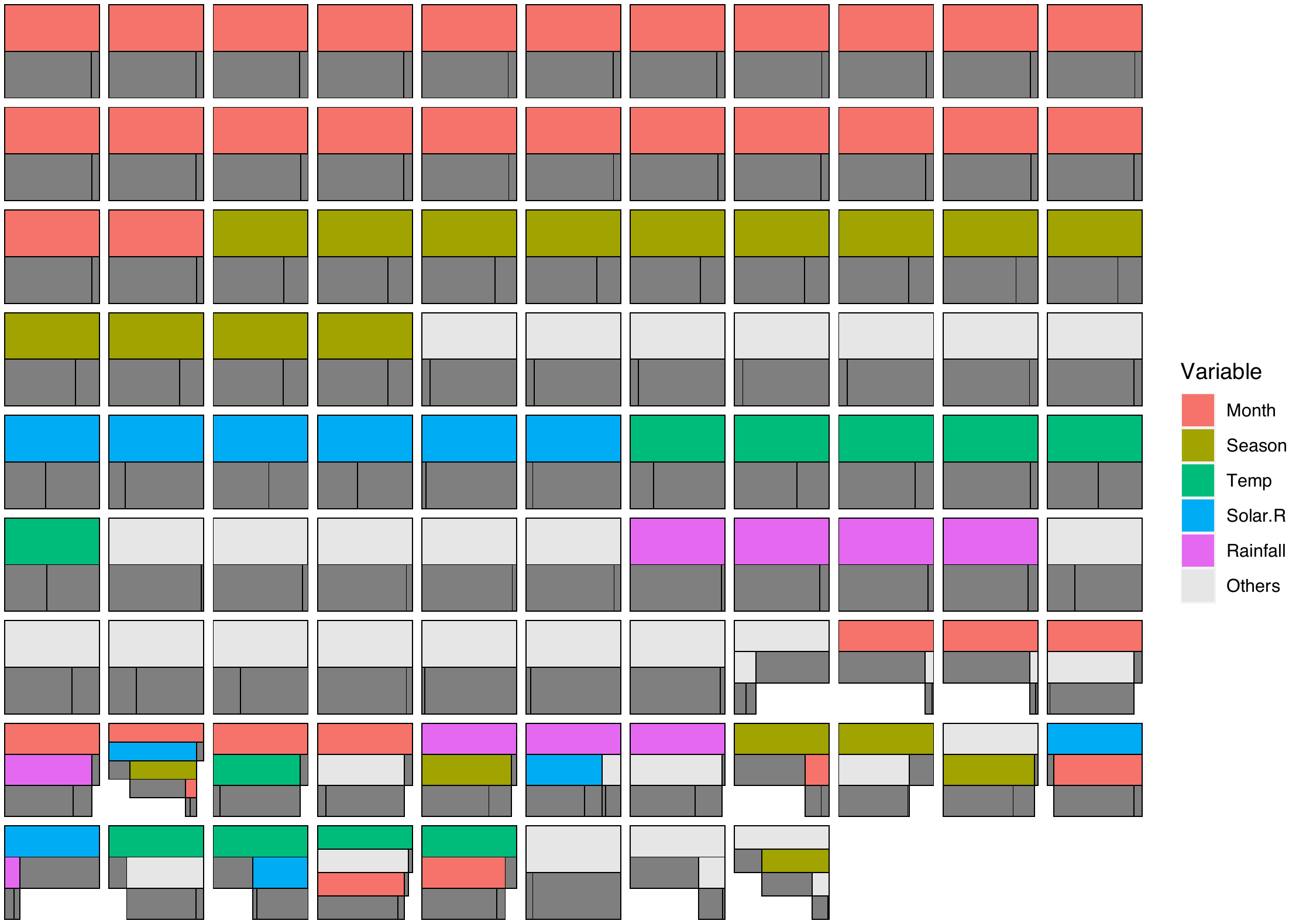}
\caption{All trees from a selected iteration, highlighting the most interesting variables and sorted by the frequency of tree type. In this case, the terminal nodes are colored dark grey and the stumps have been removed. }
\label{fig:bs_sel}
\end{figure}

In Figure \ref{fig:bs_sel} 
we take a deeper look at the structure of the trees for a selected iteration. As before we choose the iteration with the lowest residual standard deviation.  As with the importance and interactions, by default we recombine the categorical variables to display the entire factor. 
With such a large number of predictors, it can become challenging to effectively display a distinguishable hue for each when plotting the trees. To combat this we can select the most interesting variables observed in Figure \ref{fig:vivibs} and highlight them by using bright discernible colors.  To aid in efficient examination, we sort the trees by frequency of tree type and remove the stumps. 

In this iteration we can see that the most common tree is a single binary split with Month as the parent. It should also be noted that Month is chosen as the root parent more frequently than any other variable and also appears deeper in several other trees and is subsequently the most common variable found in this iteration. The previously noted interactions between Month and the other variables can be observed in the lower portion of the plot.

\begin{figure}[!t]
\centering
\includegraphics[scale = 0.4]{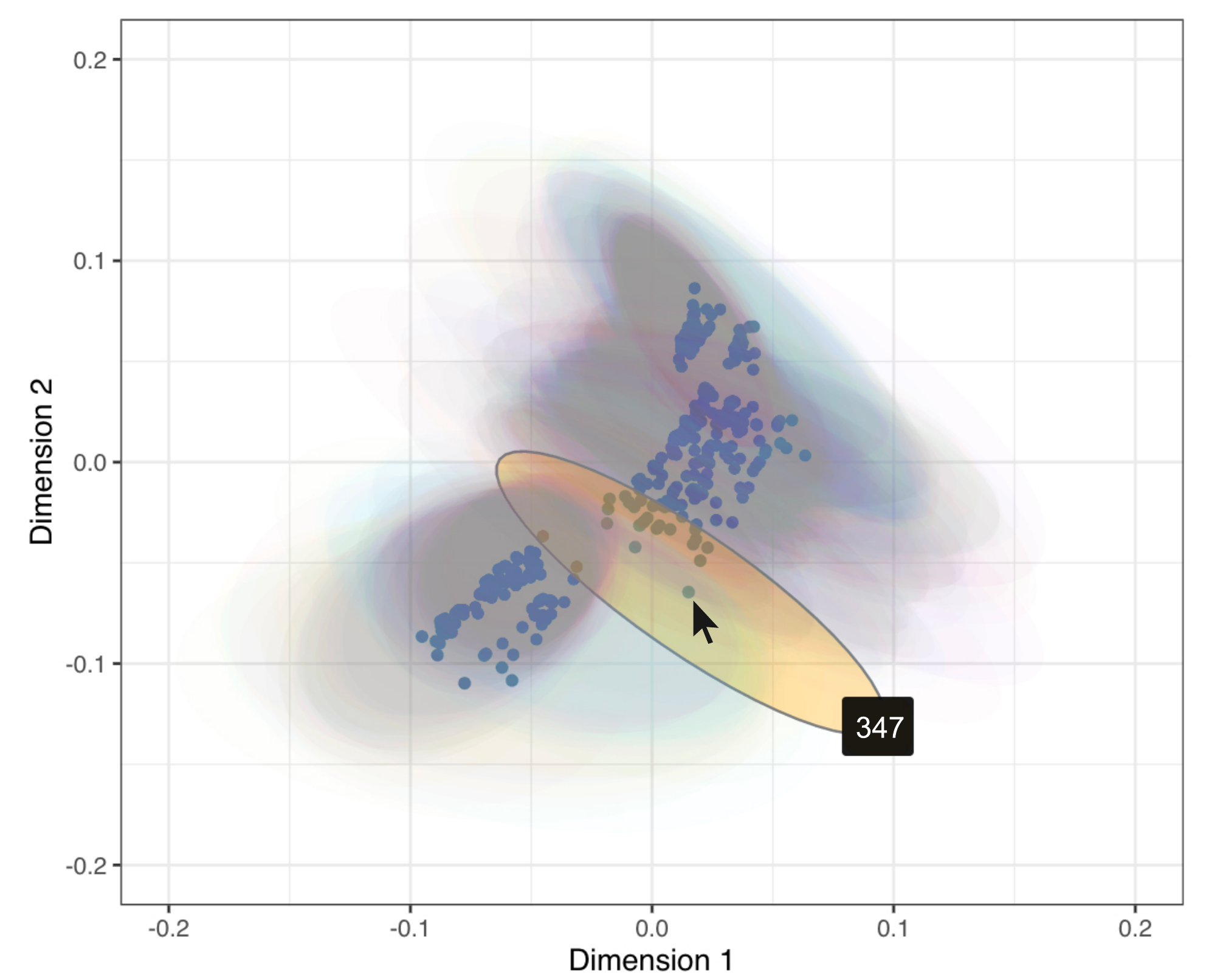}
\caption{MDS plot of a BART fit on the Seoul bike sharing data. The observations appear to have a moderate degree of uncertainty. Observation 347 (highlighted) appears to be an outlier as it lies slightly farther away from its group.}
\label{fig:bs_mds}
\end{figure}

We employ our MDS plot in Figure \ref{fig:bs_mds} 
to help find outliers. Here we can see that each observation has moderate uncertainty, represented by the surrounding 95\% uncertainty ellipses. We have highlighted observation 347 which lies slightly farther away from the group. Inspecting this observation in the data tells us that this observation corresponds to bike rentals on December $24^{th}$, which is a public holiday. Bike rentals were well below average for this day, particularly for a public holiday, which has usually high bike rentals. The temperature on this day was also well below average. This may indicate as to why this particular observation lies slightly farther from its group.

To summarize, we have used our visualizations to identify and examine  variables associated with the prediction of bike rentals in Seoul, South Korea. Our approach allowed us to examine the overall model fit and how individual and pairs of variables impact on the fit. Through our tree-based plots we can examine the inner structure of our fit. Specifically, we found the month a bike was rented was ranked the most important variable. Our methods rated Season as an important predictor, agreeing  with previous studies \citep{ve2020season}. We also find Temperature to be  important, again verifying the findings of  \cite{ve2020rule} and \cite{sathishkumar2020using}.

\section{Discussion}
\label{sec:conclusion}
We have presented new and informative visualizations for posterior evaluation of BART models. We extend the traditional method of assessing variable importance and variable interactions by
including the uncertainty that comes with Bayesian models in our point plots and heatmaps that feature the value suppressing uncertainty palettes methods of \cite{vsup}. With our tree-based plots in Section \ref{sec:treeToy} we can examine the structure of the decision trees that are created when building the model as well as providing useful summaries of tree types by way of grouping tree structures by different metrics. 
We display outlier detection methods by way of an interactive multidimensional scaling plot in Section \ref{sec:mds} to provide an in-depth examination of a model's fit. Finally, we provide a selection of enhanced model diagnostic plots in Section \ref{sec:diagPlots}, which are practical for assessing a model fit via a suite of plots that visualize aspects of a model such as stability, tree acceptance rate, average number of nodes, and average tree depth plots. These plots also provide a useful summary of the overall model fit via convergence, residual, and Q-Q plots (for regression), and ROC, precision-recall, and confusion matrices (for classification). Our approach is simple to use, adaptable,  customisable, and can be useful for comparing different BART model fits. 

Our importance and interaction plots can be useful in determining which variables have the greatest impact on the response and the inclusion of uncertainty can help in deciding if a given variable's importance is worthwhile. A drawback to this method is that the use of inclusion proportions as an importance/interaction measure relies on the splitting rules in the model. Since BART chooses the splitting rule uniformly across all variables, non-important variables can be included. This effect can be mitigated by selecting a smaller number of trees, however this may limit the predictive performance of the model, as noted by \cite{chipman2010bart}. The examples of Section \ref{sec:suppMat} show that using the proportions alone as an importance measure can be misleading, but that the use of a VSUPs with relative uncertainty provide a correction.

A current drawback occurs when the number of trees and/or MCMC iterations is large, so that the computational time to build the data frame of trees used for producing these visualizations can vary, depending on the R package used. For example, a model with 20 trees and 500 MCMC iterations took approximately 8.2, 9.2, and 90 seconds for a \texttt{BART}, \texttt{dbarts}, and \texttt{bartMachine} fit, respectively, on a MacBook Pro 2.3 GHz Dual-Core Intel Core i5 with 8GB of RAM. The disparity between \texttt{bartMachine} and the other packages is due to the way \texttt{bartMachine} uses a Java back-end to extract the raw node data from the model. Although some steps were taken to speed up this process, it remains largely outside of our control.

Our methods are flexible and can be easily extended to work with other BART packages, such as \texttt{bayesplot} \citep{bayesplot}, which is an R package that provides a large library of plotting methods for use with Bayesian models fits. Similarly, our methods could be extended to incorporate different extensions of BART, such as the methods of \cite{prado2021bayesian} for model trees BART (MOTR-BART). Rather than having a single value for the prediction at the node level, MOTR-BART estimates a linear predictor using the covariates that were used as split variables in the relevant tree.
A different method for measuring the importance and interactions could also be investigated for future work, such as DART \citep{linero2018bayesian}, which modifies a BART model by placing a Dirichlet hyper-prior on the splitting proportions of the regression tree prior. When using DART, \cite{linero2018bayesian} recommend selecting predictor variables from a so-called median probability model \citep{barbieri2004optimal} to conduct variable selection, where the median probability model is defined as a model containing variables whose posterior inclusion probability is at least 50\%. Alternatively, Shapley values \citep{shapley1997value} could be used to measure importance.






\bibliographystyle{ba}
\bibliography{bartman}


\end{document}